\documentclass[vecphys]{svmult}

\usepackage{amsmath}

\usepackage{makeidx}         
\usepackage{graphicx}        
\usepackage{multicol}        
\usepackage[bottom]{footmisc}

\makeindex             

\newcommand\beq{\begin{equation}}

\newcommand\eeq{\end{equation}}
\newcommand\beqa{\begin{eqnarray}}
\newcommand\eeqa{\end{eqnarray}}

\newcommand{\openone}{\mathsf{I}}

\newcommand{\dd}{\text{d}}

\begin{document}

\title*{Kinetic Theory for Binary Granular Mixtures at Low-Density}
\author{Vicente Garz\'{o}}
\institute{Departamento de F\'{\i}sica, Universidad de
Extremadura, E-06071 Badajoz, Spain\\ \texttt{vicenteg@unex.es}}
%
%
\maketitle

Many features of granular media can be modelled as a fluid of hard
spheres with {\em inelastic} collisions. Under rapid flow
conditions, the macroscopic behavior of grains can be described
through hydrodynamic equations. At low-density, a fundamental
basis for the derivation of the hydrodynamic equations and
explicit expressions for the transport coefficients appearing in
them is provided by the Boltzmann kinetic theory conveniently
modified to account for inelastic binary collisions. The goal of
this chapter is to give an overview of the recent advances made
for binary granular gases by using kinetic theory tools. Some of
the results presented here cover aspects such as transport
properties, energy nonequipartition, instabilities, segregation or
mixing, non-Newtonian behavior, \ldots. In addition, comparison of
the analytical results with those obtained from Monte Carlo and
molecular dynamics simulations is also carried out, showing the
reliability of kinetic theory to describe granular flows even for
strong dissipation.

\section{Introduction}
\label{sec1}

Granular systems have attracted the attention of the physics
community in the past few years, in part because the behavior of
these systems under many conditions exhibit a great similarity to
ordinary fluids \cite{H83}. These conditions include rapid, dilute
flows where the dominant transfer of momentum and energy is
through binary collisions of the grains. The main difference from
ordinary fluids is the absence of energy conservation, leading to
both obvious and subtle modifications of the usual macroscopic
balance equations as well as the constitutive equations for the
irreversible fluxes. However, in spite of the utility of
hydrodynamics to describe rapid granular flows, there are still
some open questions about its domain of validity and the
associated constitutive equations appearing in the hydrodynamic
equations \cite{K99}.

To isolate the effects of such collisional dissipation from other
important properties of granular media, an idealized microscopic
model system is usually considered: a system composed by smooth
hard spheres with inelastic collisions. As in the elastic case,
the collisions are specified in terms of the change in relative
velocity at contact but with a decrease in the magnitude of the
normal component measured by a positive coefficient of restitution
$\alpha \leq 1$. This parameter distinguishes the granular fluid
($\alpha <1$) from the ordinary fluid ($\alpha=1$). Given that the
hard sphere system with elastic collisions has been widely studied
for both equilibrium and non-equilibrium statistical mechanics
\cite{HM86}, it is tempting to apply the same methods for the case
of inelastic collisions. However, some care is warranted in
translating properties of ordinary fluids to granular fluids. In
this presentation, a kinetic theory description  based on the
Boltzmann kinetic equation (which applies at sufficiently low
density) will be considered as the appropriate tool to study
granular flows from a microscopic point of view.

Although many efforts have been devoted in the past few years to
the understanding of granular fluids, the derivation of the form
of the constitutive equations with explicit expressions for the
transport coefficients is still a subject of interest and
controversy.  The conditions to obtain a hydrodynamic description
are expected to be similar to those for normal fluids. For a given
initial state there are two stages of evolution. First, during the
kinetic stage there is rapid velocity relaxation to a
``universal'' velocity distribution that depends on the average
local density, temperature, and flow velocity. Subsequently, the
hydrodynamic stage is described through a slower evolution of
these local hydrodynamic fields as they approach uniformity. The
solution to the Boltzmann equation in this second stage is said to
be {\em normal}, where all space and time dependence of the
distribution function occurs through the macroscopic hydrodynamic
fields. The Chapman--Enskog method \cite{CC70} provides a
constructive means to obtain an approximation to such a solution
for states whose spatial gradients are not too large. In this
case, the explicit form of this normal solution is given as a
perturbation expansion in the spatial gradients of the fields.
This solution is then used to evaluate the fluxes in the
macroscopic balance equations in terms of these gradients. To
lowest order the balance equations become the granular Euler
equations while to second order they are the granular
Navier--Stokes equations. In carrying out this analysis, explicit
forms for the transport coefficients are obtained as functions of
the coefficient of restitution and other parameters of the
collision operator. In this general context, the study of
hydrodynamics for granular gases is the same as that for ordinary
fluids.

The derivation of hydrodynamics from the inelastic Boltzmann
equation has been widely covered in the case of a monodisperse gas
where the particles are of the same mass and size. As for elastic
collisions, the transport coefficients are given in terms of the
solutions of linear integral equations \cite {BDKS98,BC01}, which
are approximately solved by using Sonine polynomial expansions.
The estimates for the transport coefficients provided by the
Sonine solution compare in general quite well with both direct
Monte Carlo simulation (DSMC) of the Boltzmann equation and
molecular dynamics (MD) simulation of the gas, even for relatively
strong degrees of dissipation \cite{BRC99,BRCG00,BM04,MSG05}. This
good agreement supports the formal theoretical analysis and the
claim that hydrodynamics is not limited to nearly elastic
particles \cite{D99,D02,G03}.

Nevertheless, a real granular system is generally characterized by
some degrees of polydispersity in density and size, which leads to
phenomena very often observed in nature and experiments, such as
separation or segregation. Needless to say, the study of granular
mixtures is much more complicated than for a monodisperse gas
since not only the number of transport coefficients in a
multicomponent system is higher than that of a single gas, but
also they depend on parameters such as masses, sizes, composition
as well as several independent coefficients of restitution
$\alpha_{ij}$. Due to these difficulties, studies for
multicomponent gases are more scarce in the literature. Many of
the previous attempts \cite{JM89} to derive hydrodynamics from
kinetic theory were carried out in the quasi-elastic limit where
the equipartition of energy can be considered as an acceptable
assumption. In addition, according to this level of approximation,
the inelasticity is only accounted for by the presence of a sink
term in the energy balance equation, so that the expressions for
the transport coefficients are the same as those obtained for
ordinary fluids. However, the theoretical prediction of the
failure of energy equipartition in multicomponent granular gases
\cite{GD99} has been confirmed by computer simulations
\cite{computer}, and even observed in real experiments \cite{exp}.

Although the possibility of nonequipartition was already pointed
out many years ago \cite{JM87}, it has not been until recently
that a systematic study of the effect of nonequipartition on the
Navier--Stokes hydrodynamic equations has been carefuly carried
out  \cite{GD02,GMD06}. These new equations and associated
transport coefficients provide a somewhat more stringent test of
the analysis since the parameter space is much larger.  As in the
monodisperse case, explicit expressions for the transport
coefficients requires also to consider Sonine polynomial
expansions. The numerical accuracy of this Sonine expansion has
been confirmed by comparison with Monte Carlo simulations of the
Boltzmann equation in the cases of the shear viscosity \cite{MG03}
and the tracer diffusion \cite{GM04} coefficients. Exceptions to
this agreement are extreme mass or size ratios and strong
dissipation, although these discrepancies between theory and
simulation diminish as one considers more terms in the Sonine
polynomial approximation \cite{GM04}.

The explicit knowledge of the Navier-Stokes transport coefficients
allows quantitative application of the nonlinear hydrodynamic
equations to a number of interesting problems for granular
mixtures, such as to quantify the violation of the Einstein
relation \cite{DG01,G04} or the Onsager reciprocal relations
\cite{GMD06}, the stability analysis of the homogeneous cooling
state \cite{GMD06}, and segregation induced by a thermal gradient
\cite{G06}. In all the cases, the analysis clearly shows the
important role played by the inelasticity in the different
physical situations.

The analogy between rapid granular flow and ordinary fluids can be
also extended to many other transport situations. A particularly
simple case, allowing detailed analysis even in far from
equilibrium conditions is the simple or uniform shear flow (USF)
problem. Macroscopically, it is characterized by uniform density
and temperature and a constant mean velocity profile. This is a
well-known non-equilibrium problem widely studied, for both
granular monodisperse \cite{G03,C90,USF} and ordinary gases
\cite{GS03}. Nevertheless, the nature of this state is quite
different in each system. While for elastic fluids the temperature
increases monotonically in time due to viscous heating, a steady
state is possible for granular media when the effect of the
viscous heating is exactly compensated by the dissipation in
collisions. Thus, in the steady state, there is an intrinsic
connection between the shear field and dissipation so that the
collisional cooling sets the strength of the velocity gradient. As
a consequence, the USF state is inherently non-Newtonian and the
rheological properties of the system cannot be obtained from the
Navier-Stokes description, at least for finite dissipation
\cite{SGD04}.

The aim of this chapter is to offer a short review of recent
results obtained for binary granular mixtures from the Boltzmann
kinetic theory. It is structured as follows. The Boltzmann kinetic
equation for a granular binary mixture and its associated
macroscopic balance equations are introduced in Sec. \ref{sec2}.
Section \ref{sec3} deals with the solutions to the Boltzmann
equation for homogeneous states in the free cooling case as well
as when the mixture is heated by an external thermostat. The
Chapman--Enskog method around the local version of the homogenous
distributions obtained in Sec. \ref{sec2} is applied in Sec.
\ref{sec3} to get the form of the Navier--Stokes hydrodynamic
equations. Theoretical results for the diffusion and shear
viscosity transport coefficients are compared with simulation data
in Sec. \ref{sec5}, while the Einstein and the Onsager relations
for granular mixtures are analyzed in Secs. \ref{sec6} and
\ref{sec7}, respectively. The dispersion relations for the
hydrodynamic equations linearized about the homogeneous cooling
state are obtained in Sec. \ref{sec8}, showing that the
homogeneous reference state is unstable to long wavelength
perturbations. The conditions for stability are identified as
functions of the wave vector, the dissipation, and the parameters
of the mixture. Segregation due to thermal diffusion is studied in
Sec. \ref{sec9} by using the Navier--Stokes description. A new
criterion for segregation is found that is consistent with recent
experimental results. Section \ref{sec10} deals with the USF
problem for a granular mixture. Finally, the paper is closed in
Sec. \ref{sec11} with a discussion of the results presented here.

Before ending this section, I want to remark that the present
account is a personal perspective based on the author's work and
that of his collaborators so that no attempt is made to include
the extensive related work of many others in this field. The
references given are selective and apologies are offered at the
outset to the many other important contributions not recognized
explicitly.

\section{Boltzmann Kinetic Equation for Binary Mixtures of Inelastic Hard Spheres}
\label{sec2}

 Consider a binary granular mixture composed by smooth inelastic disks ($d=2$) or
spheres ($d=3$) of masses $m_{1}$ and $m_2$, and diameters $\sigma
_{1}$ and $\sigma_2$. The inelasticity of collisions among all
pairs is characterized by three independent constant coefficients
of restitution $\alpha_{11}$, $\alpha_{22}$, and $\alpha
_{12}=\alpha _{21}$, where $\alpha _{ij}\leq 1$ is the coefficient
of restitution for collisions between particles of species $i$ and
$j$. Since the spheres are assumed to be perfectly smooth, only
the translational degrees of freedom of grains are affected by
dissipation. In the low density regime, a simultaneous interaction
of more than two particles is highly unlike and so can be
neglected. Consequently, in a dilute gas the interactions among
the particles reduce to a succession of {\em binary} collisions.
At this level of description, all the relevant information on the
state of the system is contained in the one-body velocity
distribution functions $f_{i}({\bf r},{\bf v};t)$ ($i=1,2$)
defined so that $f_{i}({\bf r},{\bf v};t) \dd{\bf r}\dd{\bf v}$ is
the most probable (or average) number of particles of species $i$
which at time $t$ lie in the volume element $\dd{\bf r}$ centered
at the point ${\bf r}$ and moving with velocities in the range
$\dd{\bf v}$ about ${\bf v}$. For an inelastic gas, the
distributions $f_{i}({\bf r},{\bf v};t)$ ($i=1,2$) for the two
species satisfy the coupled nonlinear Boltzmann equations
\cite{GS95,BDS97}
\begin{equation}
\left( \partial _{t}+{\bf v}\cdot\nabla+\frac{{\bf F}_i}{m_i}
\cdot \frac{\partial}{\partial {\bf v}} +{\cal F}_i\right)
f_{i}({\bf r},{\bf v},t)=\sum_{j=1}^2 J_{ij}\left[ {\bf
v}|f_{i}(t),f_{j}(t)\right] \;, \label{2.1}
\end{equation}
where the Boltzmann collision operator $J_{ij}\left[ {\bf
v}|f_{i},f_{j}\right] $ is
\begin{eqnarray}
J_{ij}\left[ {\bf v}_{1}|f_{i},f_{j}\right] &=&\sigma
_{ij}^{d-1}\int \dd{\bf v} _{2}\int \dd\widehat{\boldsymbol
{\sigma }}\,\Theta (\widehat{{\boldsymbol {\sigma }}} \cdot {\bf
g}_{12})(\widehat{\boldsymbol {\sigma }}\cdot {\bf v}_{12})
\nonumber
\\
&&\times \left[ \alpha _{ij}^{-2}f_{i}({\bf r},{\bf v}_{1}^{\prime
},t)f_{j}( {\bf r},{\bf v}_{2}^{\prime },t)-f_{i}({\bf r},{\bf v}
_{1},t)f_{j}({\bf r}, {\bf v}_{2},t)\right] \;. \label{2.2}
\end{eqnarray}
In Eq.\ (\ref{2.2}), $d$ is the dimensionality of the system,
$\sigma _{ij}=\left( \sigma _{i}+\sigma _{j}\right) /2$,
$\widehat{\boldsymbol {\sigma}}$ is an unit vector along the line
of centers, $\Theta $ is the Heaviside step function, and ${\bf
v}_{12}={\bf v}_{1}-{\bf v}_{2}$ is the relative velocity. The
primes on the velocities denote the initial values $\{{\bf
v}_{1}^{\prime }, {\bf v}_{2}^{\prime }\}$ that lead to $\{{\bf
v}_{1},{\bf v}_{2}\}$ following a binary (restituting) collision:
\begin{equation}
{\bf v}_{1}^{\prime }={\bf v}_{1}-\mu _{ji}\left( 1+\alpha
_{ij}^{-1}\right) (\widehat{{\boldsymbol {\sigma }}}\cdot {\bf
v}_{12})\widehat{{\boldsymbol {\sigma }}} ,\nonumber\\
\end{equation}
\begin{equation}
 {\bf v}_{2}^{\prime }={\bf v}_{2}+\mu _{ij}\left( 1+\alpha
_{ij}^{-1}\right) (\widehat{{\boldsymbol {\sigma }}}\cdot {\bf
v}_{12})\widehat{ \boldsymbol {\sigma}} ,  \label{2.3}
\end{equation}
where $\mu _{ij}\equiv m_{i}/\left( m_{i}+m_{j}\right) $. In
addition, ${\bf F}_i$ denotes an external {\em conservative} force
acting on species $i$ (such as a gravity field) and ${\cal F}_i$
is an operator representing a possible effect of an external {\em
nonconservative} forcing which injects energy into the system to
compensate for the energy dissipated by collisional cooling. This
type of force acts as a {\em thermostat} that tries to mimics a
thermal bath. Some explicit forms for the operator ${\cal F}_i$
will be chosen later.

The relevant hydrodynamic fields for the mixture are the number
densities $n_{i}$, the flow velocity $ {\bf u}$, and the
temperature $T$. They are defined in terms of moments of the
velocity distribution functions $f_{i}$ as
\begin{equation}
\label{2.3.1} n_{i}=\int \dd{\bf v}f_{i}({\bf v})\;,
\end{equation}
\begin{equation}
\rho {\bf u}=\sum_{i=1}^2m_{i}\int \dd {\bf v}{\bf v}f_{i}({\bf
v})\;, \label{2.4}
\end{equation}
\begin{equation}
nT=p=\sum_{i=1}^2n_iT_i=\sum_{i=1}^2\frac{m_{i}}{d}\int \dd{\bf
v}V^{2}f_{i}({\bf v})\;, \label{2.5}
\end{equation}
where ${\bf V}={\bf v}-{\bf u}$ is the peculiar velocity, $
n=n_{1}+n_{2}$ is the total number density, $\rho
=m_{1}n_{1}+m_{2}n_{2}$ is the total mass density, and $p$ is the
pressure. Furthermore, the third equality of Eq.\ (\ref{2.5})
defines the kinetic temperatures $T_i$ for each species, which
measure their mean kinetic energies.

The collision operators conserve the particle number of each
species and the total momentum, but the total energy is not
conserved:
\begin{equation}
\int \dd{\bf v}J_{ij}[{\bf v}|f_{i},f_{j}]=0\;,  \label{2.6}
\end{equation}
\begin{equation}
m_i\int \dd{\bf v}{\bf v}J_{ij}[{\bf v}|f_{i},f_{j}]+m_j\int
\dd{\bf v}{\bf v}J_{ji}[{\bf v}|f_{j},f_{i}]={\bf 0} \;,
\label{2.7}
\end{equation}
\begin{equation}
\sum_{i=1}^2\sum_{j=1}^2\;m_i\int \dd{\bf v}V^{2}J_{ij}[{\bf v}
|f_{i},f_{j}]=-d nT\zeta \;.  \label{2.8}
\end{equation}
In Eq.\ (\ref{2.8}), $\zeta$ is identified as the total ``cooling
rate'' due to collisions among all species. It measures the rate
of energy loss due to dissipation. At a kinetic level, it is also
convenient to introduce the ``cooling rates'' $\zeta_i$ for the
partial temperatures $T_i$. They are defined as
\begin{equation}
\label{2.7.1} \zeta_i=\sum_{j=1}^2\zeta_{ij}=-
\frac{m_i}{dn_iT_i}\sum_{j=1}^2\int \dd{\bf v}V^{2}J_{ij}[{\bf
v}|f_{i},f_{j}],
\end{equation}
where the second equality defines the quantities $\zeta_{ij}$. The
total cooling rate $\zeta$ can be written in terms of the partial
cooling rates $\zeta_i$ as
\begin{equation}
\label{2.7.2} \zeta=T^{-1}\sum_{i=1}^2\;x_iT_i\zeta_i,
\end{equation}
where $x_i=n_i/n$ is the mole fraction of species $i$.

From Eqs.\ (\ref{2.6})--(\ref{2.8}), the macroscopic balance
equations for the number densities $n_i$, the total momentum
density $\rho {\bf u}$ and the energy density $(d/2)nT$ can be
obtained. They are given, respectively, by \cite{GD02}
\begin{equation}
D_{t}n_{i}+n_{i}\nabla \cdot {\bf u}+\frac{\nabla \cdot {\bf
j}_{i}}{m_i} =0\;, \label{2.9}
\end{equation}
\begin{equation}
\rho D_{t}{\bf u}+\nabla \cdot {\sf P}=\sum_{i=1}^2n_i{\bf F}_i\;,
\label{2.10}
\end{equation}
\begin{equation}
D_{t}T-\frac{T}{n}\sum_{i=1}^2\frac{\nabla \cdot {\bf
j}_{i}}{m_{i}}+\frac{2}{dn} \left( \nabla \cdot {\bf q}+{\sf
P}:\nabla {\bf u}-\sum_{i=1}^2\frac{{\bf F}_i\cdot {\bf
j}_i}{m_i}\right) =-(\zeta-\xi) \,T\;. \label{2.11}
\end{equation}
In the above equations, $D_{t}=\partial _{t}+{\bf u}\cdot \nabla $
is the material derivative,
\begin{equation}
{\bf j}_{i}=m_{i}\int \dd{\bf v}\,{\bf V}\,f_{i}({\bf v})
\label{2.11bb}
\end{equation}
is the mass flux for species $i$ relative to the local flow,
\begin{equation}
{\sf P}=\sum_{i=1}^2\,m_i\,\int \dd{\bf v}\,{\bf V}{\bf
V}\,f_{i}({\bf v})  \label{2.12}
\end{equation}
is the total pressure tensor, and
\begin{equation}
{\bf q}=\sum_{i=1}^2\,\frac{m_i}{2}\int \dd{\bf v}\,V^{2}{\bf
V}\,f_{i}({\bf v})  \label{2.13}
\end{equation}
is the total heat flux. On the right-hand side of the temperature
equation (\ref{2.11}), the source term $\xi$ (measuring the rate
of heating due to the external thermostat) is given by
\begin{equation}
\label{2.14} \xi=-\frac{1}{dnT}\sum_{i=1}^2\; m_i\int \dd{\bf
v}V^{2}{\cal F}_if_{i}({\bf v}).
\end{equation}
In the balance equations (\ref{2.9})--(\ref{2.11}) it is assumed
that the external driving thermostat does not change the number of
particles of each species or the total momentum, i.e.,
\begin{equation}
\label{2.15} \int \dd{\bf v}\,{\cal F}_if_{i}({\bf v})=0,
\end{equation}
\begin{equation}
\label{2.16} \sum_{i=1}^2\;m_i\int \dd{\bf v}\,{\bf v}\;{\cal
F}_if_{i}({\bf v})={\bf 0}.
\end{equation}

The utility of the balance equations (\ref{2.9})--(\ref{2.11}) is
limited without further specification of the fluxes and the
cooling rate, which in general have a complex dependence on space
and time. However, for sufficiently large space and time scales,
one expects that the system reaches a hydrodynamic regime in which
all the space and time dependence is given entirely through a
functional dependence on the six hydrodynamic fields $n_i$, ${\bf
u}$, and $T$. The corresponding functional dependence of ${\bf
j}_i$, ${\sf P}$, and ${\bf q}$ on these fields are called
constitutive equations and define the transport coefficients of
the mixture. The primary feature of a hydrodynamic description is
the reduction of the description from many microscopic degrees of
freedom to a set of equations involving only  six local fields. At
a kinetic level, the constitutive equations are obtained when one
admits the existence of a {\em normal} solution to the Boltzmann
equation where the velocity distribution functions depend on ${\bf
r}$ and $t$ only through their functional dependence on the
fields, namely,
\begin{equation}
\label{2.15.1} f_i({\bf r}, {\bf v}_1,t)=f_i[{\bf v}_1|n_i({\bf
r},t), T({\bf r},t), {\bf u}({\bf r},t)].
\end{equation}
This normal solution is generated by the Chapman--Enskog method
\cite{CC70} conveniently adapted to dissipative dynamics. Since
the method is based on an expansion around the local version of
the homogeneous state, let us characterize it before considering
inhomogeneous solutions.

\section{Homogeneous States}
\label{sec3}

In this Section we are interested in spatially homogeneous
isotropic states. In this case, we assume that the magnitude of
the conservative external forces is at least of first order in the
spatial gradients (i.e., ${\bf F}_i={\bf 0}$), so that Eq.\
(\ref{2.1}) becomes
\begin{equation}
\label{3.1} \left(\partial_t+{\cal
F}_i\right)f_i=\sum_j\;J_{ij}[f_i,f_j].
\end{equation}
For elastic collisions ($\alpha_{ij}=1$) and in the absence of
external forcing (${\cal F}_i=0$), it is well known that the
long-time solution to (\ref{3.1}) is a Maxwellian distribution for
each species at the same temperature $T$. However, if the
particles collide inelastically ($\alpha_{ij}<1$) and ${\cal
F}_i=0$, a steady state is not possible in uniform situations
since the temperature decreases monotonically in time. This state
is usually referred to as the homogeneous cooling state (HCS). In
this situation, since $n_i$ is uniform and ${\bf u}={\bf 0}$, the
normal (hydrodynamic) solution to $f_i$ requires that all its time
dependence occurs only through the temperature $T(t)$.
Consequently, $f_i(v,t)$ must be of the form \cite{GD99}
\begin{equation}
f_{i}({\bf v},t)=n_{i}v_{0}^{-d}(t)\Phi
_{i}\left(v/v_{0}(t)\right) \;, \label{3.2}
\end{equation}
where $v_{0}(t)=\sqrt{2T(t)(m_{1}+m_{2})/\left( m_{1}m_{2}\right)}
$ is a thermal speed defined in terms of the temperature $T(t)$ of
the mixture. The balance equations (\ref{2.9})--(\ref{2.11}) to
this order become $\partial _{t}x_{i}=\partial_t{\bf u}=0$, and $
T^{-1}\partial _{t}T=-\zeta$, where the cooling rate $\zeta$ is
determined by Eq.\ (\ref{2.8}). In addition, from Eqs.\
(\ref{2.7.1}) and (\ref{3.1}) (when ${\cal F}_i=0$) one can derive
the time evolution for the temperature ratio
$\gamma=T_1(t)/T_2(t)$:
\begin{equation}
\label{3.3}
\partial_t\ln \gamma=\zeta_2-\zeta_1.
\end{equation}
The fact that the distributions $f_i$ depend on time {\em only}
through $T(t)$ necessarily implies that the temperature ratio
$\gamma$ must be independent of time and so, Eq.\ (\ref{3.3})
gives the condition
\begin{equation}
\label{3.3.1} \zeta_1=\zeta_2=\zeta.
\end{equation}
In the elastic case, where $f_i$ is a Maxwellian distribution, the
above condition yields $T_1(t)=T_2(t)=T(t)$ and the energy
equipartition applies. However, in the inelastic case, the
equality of the cooling rates leads to different values for the
partial temperatures, even if one considers the Maxwellian
approximation to $f_i$. Nevertheless, the constancy of $\gamma $
assures that the time dependence of the distributions is entirely
through $T(t)$, and in fact the partial temperatures can be
expressed in terms of the global temperature as
\begin{equation}
\label{3.3.2} T_{1}(t)=\frac{\gamma}{1+x_1(\gamma-1)}T(t),\quad
T_{2}(t)=\frac{1}{1+x_1(\gamma-1)}T(t).
\end{equation}

Just as for the single gas case \cite{GS95,NE98}, the exact form
of $\Phi_i$ has not yet been found, although a good approximation
for thermal velocities can be obtained from an expansion in Sonine
polynomials \cite{GD99}. In the leading order, $\Phi_i$ is given
by
\begin{equation}
\label{3.4} \Phi_i(v^*)\to
\left(\frac{\theta_i}{\pi}\right)^{d/2}e^{-\theta_iv^{*2}}
\left[1+\frac{c_i}{4}\left(\theta_iv^{*4}-(d+2)\theta_iv^{*2}+\frac{d(d+2)}{4}\right)
\right].
\end{equation}
Here, $v^*\equiv v/v_0$,
\begin{equation}
\label{3.4.1} \theta _{i}=\frac{m_{i}}{\gamma
_{i}}\sum_{j=1}^2m_{j}^{-1},
\end{equation}
and $\gamma _{i}=T_{i}/T$. The coefficients $c_{i}$ (which measure
the deviation of $\Phi _{i}$ from the reference Maxwellian) are
determined consistently from the Boltzmann equation. The explicit
form of the approximation (\ref{3.4}) provides detailed
predictions for the temperature ratio $T_1/T_2$ (through
calculation of the cooling rates) and for the cumulants $c_i$ as
functions of the mass ratio, size ratio, composition and
coefficients of restitution \cite{GD99}. The numerical accuracy of
this truncated Sonine expansion has been confirmed by comparison
with Monte Carlo \cite{MG02bis} and MD \cite{DHGD02} simulations.

As said in the Introduction, the existence of different
temperatures for each species has been observed in real
experiments of driven steady states. These states are achieved
from external forces that do work at the same rate as collisional
cooling. In experiments this is accomplished by vibrating the
system so that it is locally driven at walls. Far from these walls
a steady state is reached whose properties are expected to be
insensitive to the details of the driving forces. Due to the
technical difficulties involved in incorporating oscillating
boundary conditions, it is usual to introduce external forces (or
thermostats) acting locally on each particle. These forces are
represented by the operator ${\cal F}_i$ in Eq.\ (\ref{3.1}) and
depend on the state of the system. Two types of thermostats have
been usually considered in the literature. One is a deterministic
thermostat widely used in nonequilibrium MD simulations for
ordinary fluids \cite{EM90,H91}. The force is similar to a Stokes
law drag force, linear in the velocity, but with the opposite sign
so that it heats rather than cools the system. In this case,
${\cal F}_i$ is given by \cite{MG03,MS00}
\begin{equation}
\label{3.5} {\cal F}_if_i({\bf
v})=\frac{1}{2}\zeta_i\frac{\partial}{\partial {\bf v}}\cdot [{\bf
v}f_i({\bf v})],
\end{equation}
where the friction constant in the force has been adjusted to get
a constant temperature in the long-time limit. It must be remarked
that the corresponding Boltzmann equation (\ref{3.1}) for this
Gaussian thermostat force is formally identical with the Boltzmann
equation in the HCS (i.e., with ${\cal F}_i=0$) when both
equations are written in terms of the reduced distributions
$\Phi_i(v^*)$ \cite{MS00}. In particular, the dependence of
$\gamma$ on the parameters of the system is the same with and
without the Gaussian thermostat.

A second method of driving the system is by means of a stochastic
Langevin force representing Gaussian white noise \cite{WM96}. The
corresponding operator ${\cal F}_i$ has a Fokker--Planck form
\cite{NE98}
\begin{equation}
\label{3.6} {\cal F}_if_i({\bf
v})=-\frac{1}{2}\frac{T_i}{m_i}\zeta_i\left(\frac{\partial}{\partial
{\bf v}}\right)^2f_i({\bf v}),
\end{equation}
where for simplicity the covariance of the stochastic acceleration
has been taken to the same for each species \cite{HBB00,BT02}.
This requirement gives the steady state condition
\begin{equation}
\label{3.7} \frac{T_1}{m_1}\zeta_1=\frac{T_2}{m_2}\zeta_2.
\end{equation}
The cooling rates $\zeta_i$ are no longer equal, in contrast to
the HCS, and the dependence of $\gamma$ on the control parameters
is different as well \cite{DHGD02}. The procedure for determining
the temperature ratio and the cumulants $c_i$ is the same as in
the HCS state since the steady state distribution $\Phi_i$ can
also be represented as an expansion of the form (\ref{3.4}) and
the coefficients are now determined from the solution to the
Boltzmann equation (\ref{3.1}). The condition (\ref{3.7}) gives
the corresponding equation for the temperature ratio.

\begin{figure}
\centering
\includegraphics[width=0.7 \columnwidth,angle=0]{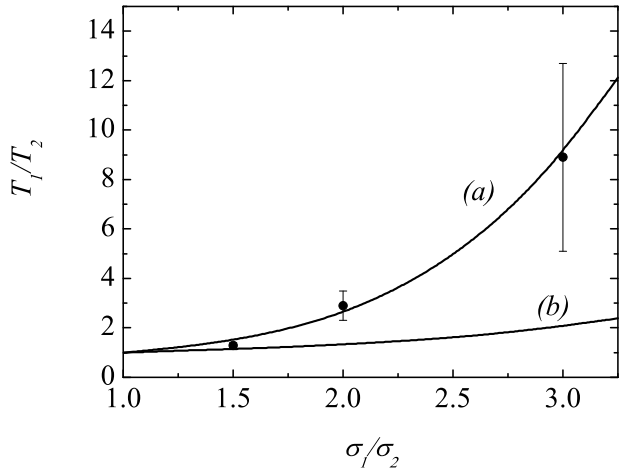}
\caption{Temperature ratio $T_1/T_2$ versus the size ratio
$\sigma_1/\sigma_2$ for $\alpha_{ij}\equiv \alpha=0.78$ in the
case of mixtures constituted by particles of the same material and
equal total volumes of large and small particles. The lines are
the kinetic theory results in (a) the stochastic driving case and
(b) the free cooling case, while the points refer to MD
simulations \cite{SUKSS06}. \label{fig1}}
\end{figure}

Figure \ref{fig1} illustrates the differences between the HCS and
the stochastic steady state at the level of the temperature ratio
$T_1/T_2$. We have considered mixtures constituted by spheres
($d=3$) of the same material [and so, $\alpha_{ij}=\alpha$, and
$m_1/m_2=(\sigma_1/\sigma_2)^3$] and equal volumes of large and
small particles [i.e., $x_2=(\sigma_1/\sigma_2)^3x_1$]. Here, for
the sake of simplicity, the cooling rates have been analytically
estimated by using Maxwellians (namely, by taking $c_i=0$) for the
distributions $f_i({\bf v})$:
\begin{eqnarray}
\label{3.7.1} \zeta_i=\sum_{j=1}^2\zeta_{ij}&\to&
\frac{4\pi^{(d-1)/2}}{d\Gamma\left(\frac{d}{2}\right)}
v_0\sum_{j=1}^2
n_j\mu_{ji}\sigma_{ij}^{d-1}\left(\frac{\theta_i+\theta_j}
{\theta_i\theta_j}\right)^{1/2}\nonumber\\
& &\times (1+\alpha_{ij})
\left[1-\frac{\mu_{ji}}{2}(1+\alpha_{ij})
\frac{\theta_i+\theta_j}{\theta_j}\right].
\end{eqnarray}
Simulation data recently obtained from MD simulations in agitated
mixtures have also been included \cite{SUKSS06}. The experimental
value of the (common) coefficient of restitution is $\alpha=0.78$.
While a good agreement between kinetic theory and MD simulations
is found when the gas is assumed to be driven by the stochastic
thermostat, significant discrepancies appear in the undriven (HCS)
case, especially as the size ratio $\sigma_1/\sigma_2$ increases.
These results contrast with the comparison made by Brey {\em et
al.} \cite{BRM05} for agitated mixtures in the tracer limit
($x_1\to 0$), where the predictions of the temperature ratio from
kinetic theory based on the condition $\zeta_1=\zeta_2$ compare
quite well with MD simulations. However, for the cases studied in
Ref.\ \cite{BRM05} the conditions (\ref{3.3.1}) and (\ref{3.7})
yield quite similar results for the dependence of $T_1/T_2$ on the
parameters of the system. The good agreement found in Fig.\
\ref{fig1} between MD simulations for agitated mixtures and
kinetic theory suggests that the stochastic driving condition can
be considered as a plausible first approximation for qualitative
comparisons with experimental results \cite{exp}.

\section{Navier--Stokes Hydrodynamic Equations}
\label{sec4}

We consider now a spatially inhomogeneous state created either by
initial preparation or by boundary conditions. We are interested
in a hydrodynamic description where the state of the system is
completely specified through their hydrodynamic fields. This
implies that the latter dominate over other excitations for times
large compared to the mean free time and for wavelengths large
compared to the mean free time. The hydrodynamic regime is
characterized by the existence of a {\em normal} solution to the
Boltzmann equation which can be explicitly obtained by means of
the Chapman--Enskog method \cite{CC70}. For small spatial
variations, the functional dependence (\ref{2.15.1}) of the normal
solution can be made local in space and time through an expansion
in gradients of the fields:
\begin{equation}
\label{4.1} f_i=f_i^{(0)}+\epsilon\; f_i^{(1)}+\cdots,
\end{equation}
where each factor of $\epsilon$ means an implicit gradient of a
hydrodynamic field. The reference distribution function
$f_i^{(0)}({\bf r}, {\bf v}, t)$ is the {\em local} version of the
homogeneous distribution (\ref{3.2}), namely, it is obtained from
the homogeneous distribution by replacing the temperature,
densities and flow velocity by their nonequilibrium values:
\begin{equation}
\label{4.2} f_i^{(0)}({\bf r}, {\bf v}, t)=n_{i}({\bf r},
t)v_{0}^{-d}(T({\bf r}, t))\Phi _{i}\left(V/v_{0}(T({\bf r},
t))\right) \;,
\end{equation}
where ${\bf V}={\bf v}-{\bf u}({\bf r}, t)$. The time derivatives
of the fields are also expanded as
$\partial_t=\partial_t^{(0)}+\epsilon \partial_t^{(1)}+\cdots$.
The coefficients of the time derivative expansion are identified
from the balance equations (\ref{2.9})--(\ref{2.11}) with a
representation of the fluxes and the cooling rate in the
macroscopic balance equations as a similar series through their
definitions as functionals of $f_i$.

This is the usual Chapman--Enkog method for solving kinetic
equations \cite{CC70,GS03,FK72}. Nevertheless, the complexity
introduced by the energy dissipation in collisions has led to the
introduction by some authors \cite{JM89} of some additional
approximations, restricting the validity of most of the results to
the small inelasticity limit. Only very recently, explicit
expressions for the fluxes to first order in the gradients as
explicit functions of the coefficients of restitution have been
obtained \cite{GD02, MG03,GM06}. To Navier--Stokes order, the
constitutive equations for mass, momentum, and heat fluxes are
given, respectively, by
\begin{equation}
{\bf j}_{1}^{(1)}=-\frac{m_{1}m_{2}n}{\rho } D\nabla x_{1}-\frac{
\rho }{p}D_{p}\nabla p-\frac{\rho }{T}D^{\prime }\nabla
T+\sum_{i=1}^2\chi_{1i}{\bf F}_i,\hspace{0.3in}{\bf
j}_{2}^{(1)}=-{\bf j}_{1}^{(1)}, \label{4.3}
\end{equation}
\begin{equation}
P_{k \ell}^{(1)}=p\;\delta _{k\ell}-\eta \left( \nabla _{\ell
}u_{k}+\nabla _{k}u_{\ell}-\frac{2}{d}\delta _{k\ell} {\bf \nabla
\cdot u}\right),  \label{4.4}
\end{equation}
\begin{equation}
{\bf q}^{(1)}=-T^{2}D^{\prime \prime }\nabla x_{1}-L\nabla
p-\lambda \nabla T+\sum_{i=1}^2 \kappa_i {\bf F}_i. \label{4.5}
\end{equation}
The transport coefficients in these equations are the diffusion
coefficient $D$, the pressure diffusion coefficient $D_p$, the
thermal diffusion coefficient $D'$, the mobility coefficient
$\chi_{ij}$, the shear viscosity $\eta$, the Dufour coefficient
$D''$, the pressure energy coefficient $L$, the thermal
conductivity $\lambda$, and the coefficient $\kappa_i$. These
coefficients are defined as
\begin{equation}
D=-\frac{\rho }{dm_{2}n}\int \dd{\bf v}\,{\bf V}\cdot {\boldsymbol
{\cal A}}_{1}, \label{4.6}
\end{equation}
\begin{equation}
D_{p}=-\frac{m_{1}p}{d\rho}\int \dd{\bf v}\,{\bf V}\cdot
{\boldsymbol {\cal B}}_{1}, \label{4.7}
\end{equation}
\begin{equation}
D^{\prime }=-\frac{m_{1}T}{d\rho }\int \dd{\bf v}\,{\bf V}\cdot
{\boldsymbol {\cal C}}_{1}, \label{4.8}
\end{equation}
\begin{equation}
\chi_{ij}=\frac{1}{d}\int \dd{\bf v}\,{\bf V}\cdot {\boldsymbol
{\cal E}}_{ij}, \label{4.8.1}
\end{equation}
\begin{equation}
\eta =-\frac{1}{(d-1)(d+2)}\sum_{i=1}^2\,m_i\,\int \dd{\bf v}\,
{\bf V}{\bf V}:{\boldsymbol {\cal D}}_{i}, \label{4.9}
\end{equation}
\begin{equation}
D^{\prime \prime
}=-\frac{1}{dT^{2}}\sum_{i=1}^2\,\frac{m_i}{2}\,\int \dd{\bf
v}\,V^{2}{\bf V}\cdot {\boldsymbol {\cal A}}_{i}, \label{4.10}
\end{equation}
\begin{equation}
L=-\frac{1}{d}\sum_{i=1}^2\,\frac{m_i}{2}\,\int \dd{\bf
v}\,V^{2}{\bf V} \cdot \,{\boldsymbol {\cal B}}_{i}, \label{4.11}
\end{equation}
\begin{equation}
\lambda =-\frac{1}{d}\sum_{i=1}^2\,\frac{m_i}{2}\,\int \dd{\bf
v}\,V^{2} {\bf V}\cdot {\boldsymbol {\cal C}}_{i}, \label{4.12}
\end{equation}
\begin{equation}
\kappa_i =\frac{1}{d}\sum_{j=1}^2\,\frac{m_i}{2}\,\int \dd{\bf
v}\,V^{2} {\bf V}\cdot {\boldsymbol {\cal E}}_{ij}. \label{4.12.1}
\end{equation}
Here, ${\boldsymbol {\cal A}}_{i}({\bf V})$, ${\boldsymbol {\cal
B}}_{i}({\bf V})$, ${\boldsymbol {\cal C}}_{i}({\bf V})$,
${\boldsymbol {\cal D}}_{i}({\bf V})$, and
 ${\boldsymbol {\cal E}}_{ij}({\bf V})$ are
functions of the peculiar velocity and the hydrodynamic fields.
They obey the following set of coupled linear integral equations:
\begin{subequations}
\begin{eqnarray}
\left[ -c_\zeta^{(0)}\left( T\partial _{T}+p\partial _{p}\right)
+{\cal F}_1^{(0)}+{\cal L}_{1} \right] {\boldsymbol {\cal
A}}_{1}+{\cal M}_{1}{\boldsymbol {\cal A}}_{2}&=&{\bf A}_{1}
+\left( \frac{\partial c_\zeta^{(0)}}{\partial x_{1}}\right)
_{p,T}\nonumber\\
& & \times \left( p{\boldsymbol {\cal
 B}}_{1}+T{\boldsymbol {\cal C}}_{1}\right) ,  \label{4.13.1}
\end{eqnarray}
\begin{eqnarray}
\left[ -c_\zeta^{(0)}\left( T\partial _{T}+p\partial _{p}\right)
+{\cal F}_2^{(0)}+{\cal L}_{2} \right] {\boldsymbol {\cal
A}}_{2}+{\cal M}_{2}{\boldsymbol {\cal A}}_{1}&=&{\bf
A}_{2}+\left(
\frac{\partial c_\zeta^{(0)}}{\partial x_{1}}\right) _{p,T}\nonumber\\
& & \times\left( p{\boldsymbol {\cal
 B}}_{2}+T{\boldsymbol {\cal C}}_{2}\right) ,  \label{4.13.2}
\end{eqnarray}
\end{subequations}
\begin{subequations}
\begin{equation}
\left[ -c_\zeta^{(0)}\left( T\partial _{T}+p\partial _{p}\right)
+{\cal F}_1^{(0)}+{\cal L}_{1}-2c_\zeta^{(0)}\right] {\boldsymbol
{\cal B}}_{1}+{\cal M}_{1}{\boldsymbol {\cal B}}_{2}= {\bf
B}_{1}+\frac{Tc_\zeta^{(0)}}{p}{\boldsymbol {\cal C}}_{1},
\label{4.14.1}
\end{equation}
\begin{equation}
\left[ -c_\zeta^{(0)}\left( T\partial _{T}+p\partial _{p}\right)
+{\cal F}_2^{(0)}+{\cal L}_{2}-2c_\zeta^{(0)}\right] {\boldsymbol
{\cal B}}_{2}+{\cal M}_{2}{\boldsymbol {\cal B}}_{1}= {\bf
B}_{2}+\frac{Tc_\zeta^{(0)}}{p}{\boldsymbol {\cal C}}_{2},
\label{4.14.2}
\end{equation}
\end{subequations}
\begin{subequations}
\begin{equation}
\left[ -c_\zeta^{(0)}\left( T\partial _{T}+p\partial _{p}\right)
+{\cal F}_1^{(0)}+{\cal L}_{1}-\frac{1}{2}c_\zeta^{(0)}\right]
{\boldsymbol {\cal C}}_{1}+{\cal M}_{1} {\boldsymbol {\cal
C}}_{2}= {\bf C}_{1}-\frac{pc_\zeta^{(0)}}{2T}{\boldsymbol {\cal
B}}_{1}, \label{4.15.1}
\end{equation}
\begin{equation}
\left[ -c_\zeta^{(0)}\left( T\partial _{T}+p\partial _{p}\right)
+{\cal F}_2^{(0)}+{\cal L}_{2}-\frac{1}{2}c_\zeta^{(0)}\right]
{\boldsymbol {\cal C}}_{2}+{\cal M}_{2} {\boldsymbol {\cal
C}}_{1}= {\bf C}_{2}-\frac{pc_\zeta^{(0)}}{2T}{\boldsymbol {\cal
B}}_{2}, \label{4.15.2}
\end{equation}
\end{subequations}
\begin{subequations}
\begin{equation}
\label{4.16.1} \left[ -c_\zeta^{(0)}\left( T\partial
_{T}+p\partial _{p}\right) +{\cal F}_1^{(0)}+{\cal L}_{1}\right]
{\boldsymbol {\cal D}}_{1} +{\cal M}_{1}{\boldsymbol {\cal
D}}_{2}={\sf D}_1,
\end{equation}
\begin{equation}
\label{4.16.2} \left[ -c_\zeta^{(0)}\left( T\partial
_{T}+p\partial _{p}\right) +{\cal F}_2^{(0)}+{\cal L}_{2}\right]
{\boldsymbol {\cal D}}_{2} +{\cal M}_{2}{\boldsymbol {\cal
D}}_{1}={\sf D}_2,
\end{equation}
\end{subequations}
\begin{subequations}
\begin{equation}
\label{4.n1} \left[ -c_\zeta^{(0)}\left( T\partial _{T}+p\partial
_{p}\right) +{\cal F}_1^{(0)}+{\cal L}_{1}\right] {\boldsymbol
{\cal E}}_{11} +{\cal M}_{1}{\boldsymbol {\cal E}}_{21}={\bf
E}_{11},
\end{equation}
\begin{equation}
\label{4.n2} \left[ -c_\zeta^{(0)}\left( T\partial _{T}+p\partial
_{p}\right) +{\cal F}_1^{(0)}+{\cal L}_{1}\right] {\boldsymbol
{\cal E}}_{12} +{\cal M}_{1}{\boldsymbol {\cal E}}_{22}={\bf
E}_{12}.
\end{equation}
\end{subequations}
Here, we have introduced the linearized Boltzmann collision
operators
\begin{equation}
{\cal L}_{1}X=-\left( J_{11}[f_{1}^{(0)},X]+J_{11}[X,f_{1}^{(0)}]+
J_{12}[X,f_{2}^{(0)}]\right) \;, \label{4.n3}
\end{equation}
\begin{equation}
{\cal M}_{1}X=-J_{12}[f_{1}^{(0)},X], \label{4.n4}
\end{equation}
with a similar definition for ${\cal L}_{2}$ and ${\cal M}_{2}$.
In addition, $c_\zeta^{(0)}=\zeta^{(0)}$ in the undriven case
while $c_\zeta^{(0)}=0$ in the driven case, where $\zeta^{(0)}$ is
given by Eq.\ (\ref{2.8}) to zeroth order, i.e.,
\begin{equation}
\zeta^{(0)}=-\frac{1}{d nT}\sum_{i=1}^2\sum_{j=1}^2\;m_i\int
\dd{\bf v}V^{2}J_{ij}[{\bf v} |f_{i}^{(0)},f_{j}^{(0)}]\;.
\label{4.n5}
\end{equation}
In Eqs.\ (\ref{4.13.1})--(\ref{4.n2}) we have also introduced the
operators
\begin{subequations}
\begin{equation}
\label{4.16.3} {\cal
F}_i^{(0)}X=\frac{1}{2}\zeta_i^{(0)}\frac{\partial}{\partial {\bf
V}}\cdot \left({\bf V}X\right), \quad (\text{Gaussian
thermostat}),
\end{equation}
\begin{equation}
\label{4.16.4} {\cal
F}_i^{(0)}X=-\frac{1}{2}\frac{T_i}{m_i}\zeta_i^{(0)}\left(\frac{\partial}{\partial
{\bf V}}\right)^2X, \quad (\text{stochastic thermostat}),
\end{equation}
\end{subequations}
and the quantities
\begin{equation}
{\bf A}_{i}({\bf V})=-\left(\frac{\partial}{\partial x_{1}}
f_{i}^{(0)}\right)_{p,T}{\bf V}, \label{4.17}
\end{equation}
\begin{equation}
{\bf B}_{i}({\bf V})=-\frac{1}{p}\left[ f_{i}^{(0)}{\bf
V}+\frac{nT}{\rho } \left(\frac{\partial}{\partial {\bf
V}}f_{i}^{(0)}\right) \right] , \label{4.18}
\end{equation}
\begin{equation}
{\bf C}_{i}({\bf V})=\frac{1}{T}\left[
f_{i}^{(0)}+\frac{1}{2}\frac{\partial }{\partial {\bf V}}\cdot
\left( {\bf V}f_{i}^{(0)}\right) \right] {\bf V}, \label{4.19}
\end{equation}
\begin{equation}
{\sf D}_{i}({\bf V})={\bf V}\frac{\partial }{\partial {\bf V}}
f_{i}^{(0)}-\frac{1}{d}\left({\bf V}\cdot \frac{\partial
}{\partial {\bf V}}f_{i}^{(0)}\right)\openone ,  \label{4.20}
\end{equation}
\begin{equation}
{\bf E}_{ij}({\bf V})=-\left(\frac{\partial}{\partial {\bf
V}}f_{i}^{(0)}\right)\frac{1}{m_j}\left(\delta_{ij}-\frac{\rho_j}{\rho}\right).
\label{4.20.1}
\end{equation}
Here, $\openone$ is the unit tensor in $d$ dimensions and
$\rho_i=m_in_i$ is the mass density of species $i$. Upon writing
Eqs.\ (\ref{4.13.1})--(\ref{4.n2}) use has been made of the fact
that there is no contribution to the cooling rate at this order,
i.e., $\zeta^{(1)}=0$. As a consequence, ${\cal F}_i^{(1)}=0$. The
property $\zeta^{(1)}=0$ is special of the low density Boltzmann
kinetic theory (since $f_i^{(1)}$ does not contain any
contribution proportional to $\nabla \cdot {\bf u}$), but such
terms occur at higher densities \cite{GD99b,L05}. Note that in the
particular case of the gravitational force ${\bf F}_i=m_i{\bf g}$
(where ${\bf g}$ is the gravity acceleration), the combination
$m_1{\bf E}_{11}+m_2{\bf E}_{12}={\bf 0}$. This leads to
${\boldsymbol {\cal E}}_{ij}={\bf 0}$, and so there are no
contributions to the mass and heat fluxes coming from the external
conservative forces.

In summary, the solutions to the Boltzmann equations to first
order in the spatial gradients are given by \cite{GD02}
\begin{equation}
\label{4.21} f_i=f_i^{(0)}+{\boldsymbol {\cal A}}_{i}\cdot \nabla
x_1+{\boldsymbol {\cal B}}_{i}\cdot \nabla p+{\boldsymbol {\cal
C}}_{i}\cdot \nabla T+{\boldsymbol {\cal D}}_{i}:\nabla {\bf
u}+\sum_{j}\;{\boldsymbol {\cal E}}_{ij}\cdot {\bf F}_j.
\end{equation}
The solution to zeroth-order is obtained from Eq.\ (\ref{3.1})
while the functions $\{{\boldsymbol {\cal A}}_{i},{\boldsymbol
{\cal B}}_{i},{\boldsymbol {\cal C}}_{i},{\boldsymbol {\cal
D}}_{i}, {\boldsymbol {\cal E}}_{ij}\}$ are determined from the
integral equations (\ref{4.13.1})--(\ref{4.n2}). Once these
equations are solved, the Navier-Stokes transport coefficients are
obtained from Eqs.\ (\ref{4.6})--(\ref{4.12.1}) and the mass,
momentum, and heat fluxes are explicitly known. These fluxes,
together with the macroscopic balance equations
(\ref{2.9})--(\ref{2.11}), provide the closed set of Navier-Stokes
order hydrodynamic equations for a granular binary mixture. All
these results are still formally exact and valid for arbitrary
values of the coefficients of restitution.

However, explicit expressions for the Navier--Stokes transport
coefficients require to solve the integral equations
(\ref{4.13.1})--(\ref{4.n2}). Accurate approximations for
$\{{\boldsymbol {\cal A}}_{i},{\boldsymbol {\cal
B}}_{i},{\boldsymbol {\cal C}}_{i},{\boldsymbol {\cal D}}_{i},
{\boldsymbol {\cal E}}_{ij}\}$ may be obtained using low order
truncation of expansions in a series of Sonine polynomials. The
polynomials are defined with respect to a Gaussian weight factor
whose parameters are chosen such that the leading term in the
expansion yields the exact moments of the entire distribution with
respect to $1$, ${\bf v}$, and $v^2$. The procedure is similar to
the one followed for elastic collisions \cite{CC70} and yields
explicit expressions for the transport coefficients in terms of
the parameters of the mixture \cite{GD02,GMD06,GM06}.

\section{Comparison with Monte Carlo Simulations}
\label{sec5}

As said before, the expressions derived for the Navier Stokes
transport coefficients are obtained by considering two different
approximations. First, since the deviation of $f_i^{(0)}$ from its
Maxwellian form is quite small in the region of thermal
velocities, one uses the distribution (\ref{3.4}) as a trial
function for $f_i^{(0)}$. Second, one only considers the leading
terms of an expansion of the distribution $f_i^{(1)}$ in Sonine
polynomials. Both approximations allow one to offer a simplified
kinetic theory for a granular binary mixture. To assess the degree
of accuracy of these predictions, one resorts to numerical
solutions of the Boltzmann equation, such as those obtained from
the Direct simulation Monte Carlo (DSMC) method \cite{Bird}.
Although the method was originally devised for normal fluids, its
extension to granular gases is straightforward \cite{MS00}. In
this Section we provide some comparisons between theory and
numerical solutions of the Boltzmann equation by means of the DSMC
method in the cases of the diffusion coefficient $D$ (in the
tracer limit) and the shear viscosity coefficient $\eta$ of a
heated gas. Let us study each coefficient separately.

\subsection{Tracer Diffusion Coefficient}
\label{sec5.1}

We consider a free granular mixture (${\cal F}_i=0$) in which one
of the components of the mixture (say, for instance, species $1$)
is present in tracer concentration ($x_1\to 0$). In this situation
the diffusion coefficient of impurities in a granular gas
undergoing homogeneous cooling state can be measured in
simulations from the mean square displacement of the tracer
particle after a time interval $t$ \cite{M89}:
\begin{equation}
\label{5.1} \frac{\partial}{\partial t}\langle |{\bf r}(t)-{\bf
r}(0)|^2 \rangle =\frac{2dD}{n_2},
\end{equation}
where $|{\bf r}(t)-{\bf r}(0)|$ is the distance travelled by the
impurity from $t=0$ until time $t$. The relation (\ref{5.1})
written in appropriate dimensionless variables to eliminate the
time dependence of $D(t)$ can be used to measure by computer
simulations the diffusion coefficient \cite {BRCG00,GM04}.

If the hydrodynamic description (or normal solution in the context
of the Chapman--Enskog method) applies, then the diffusion
coefficient $D(t)$ depends on time only through its dependence on
the temperature $T(t)$. Dimensional analysis shows that
$D(t)\propto \sqrt{T(t)}$. In this case, after a transient regime,
the reduced diffusion coefficient
$D^*=(m_1m_2/\rho)D(t)\nu_0(t)/T(t)$ achieves a time-independent
value. Here, $\nu_0(t)=n\sigma_{12}^{d-1}v_0(t)\propto
\sqrt{T(t)}$ is an effective collision frequency for hard spheres.
The fact that $D^*$ reaches a constant value for times large
compared with the mean free path is closely related with the
validity of a hydrodynamic description for the system. In
addition, as has been recently shown \cite{SD06}, the dependence
of $D^*$ on the mass ratio $m_1/m_2$ and the coefficient of
restitution $\alpha_{12}$ is only through the effective mass
$m_1^*=m_1+(m_1+m_2)(1-\alpha_{12})/(1+\alpha_{12})$.

\begin{figure}
\centering
\includegraphics[width=0.7 \columnwidth,angle=0]{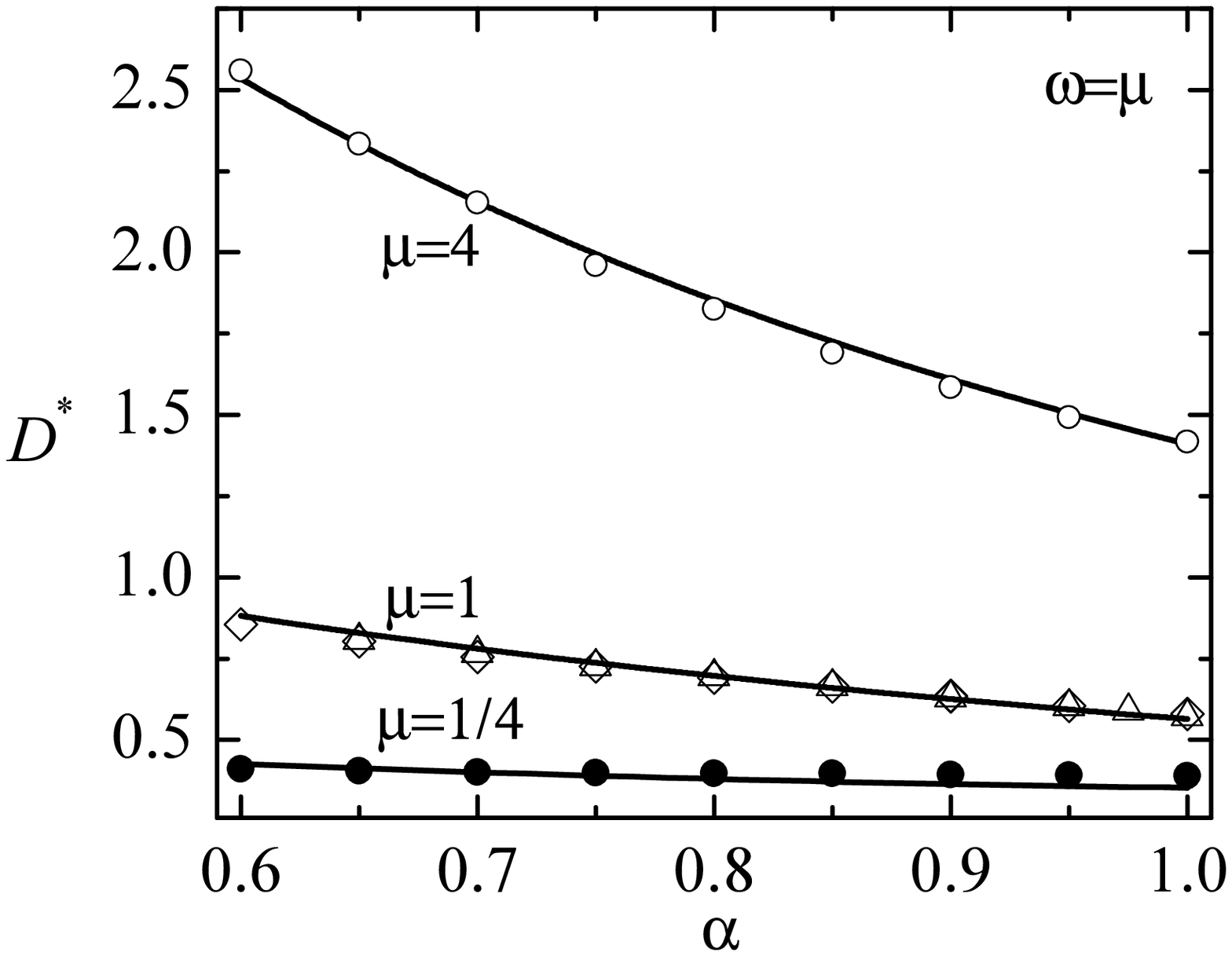}
\caption{Plot of the reduced diffusion coefficient $D^*$ as a
function of the (common) coefficient of restitution $\alpha$ for
binary mixtures with $\omega=\mu$ in the case of a two-dimensional
system ($d=2$). Here, $\omega \equiv \sigma_1/\sigma_2$ and $\mu
\equiv m_1/m_2$. The symbols are computer simulation results
obtained from the mean square displacement and the lines are the
theoretical results obtained in the first Sonine approximation.
The DSMC results correspond to $\mu=1/4$ ({\Large $\bullet$}),
$\mu=4$ ({\Large $\circ$}) and $\mu=1$ ({\Large $\diamond$}).
Molecular dynamics results reported in Ref.\ \cite{BRCG00} for
$\mu=1$ ($\triangle$) have also been included.} \label{fig2}
\end{figure}

The dependence of $D^*$ on the common coefficient of restitution
$\alpha_{ij}\equiv \alpha$ is shown in Fig.\ \ref{fig2} in the
case of hard disks ($d=2$) for three different systems. The
symbols refer to DSMC simulations while the lines correspond to
the kinetic theory results obtained in the first Sonine
approximation \cite{GD02,GM06}. MD results reported in Ref.\
\cite{BRCG00} when impurities and particles of the gas are
mechanically equivalent have also been included. We observe that
in the latter case MD and DSMC results are consistent among
themselves in the range of values of $\alpha$ explored. This good
agreement gives support to the applicability of the inelastic
Boltzmann equation beyond the quasielastic limit. It is apparent
that the agreement between the first Sonine approximation and
simulation results is excellent when impurities and particles of
the gas are mechanically equivalent and when impurities are much
heavier and/or much larger than the particles of the gas (Brownian
limit). However, some discrepancies between simulation an theory
are found with decreasing values of the mass ratio $\mu \equiv
m_1/m_2$ and the size ratio $\omega \equiv \sigma_1/\sigma_2$.
These discrepancies are not easily observed in Fig.\ \ref{fig2}
because of the small magnitude of $D^*$ for $\mu=1/4$. For these
systems, the second Sonine approximation \cite{GM04} improves the
qualitative predictions over the first Sonine approximation for
the cases in which the gas particles are heavier and/or larger
than impurities. This means that the Sonine polynomial expansion
exhibits a slow convergence for sufficiently small values of the
mass ratio $\mu$ and/or the size ratio $\omega$. This tendency is
also present in the case of elastic systems \cite{MC84}.

\subsection{Shear Viscosity Coefficient of a Heated Gas}

The shear viscosity $\eta$ is perhaps the most widely studied
transport coefficient in granular fluids. This coefficient can be
measured in computer simulations in the special hydrodynamic state
of uniform shear flow (USF). At a macroscopic level, this state is
characterized by constant partial densities $n_i$, uniform
temperature $T$, and a linear flow velocity profile $u_{1,k}=
u_{2,k}=a_{k\ell}r_\ell$, $a_{k\ell}=a\delta_{kx}\delta_{\ell y}$,
$a$ being the constant shear rate. In this state, the temperature
changes in time due to the competition between two mechanisms: on
the one hand, viscous heating and, on the other hand, energy
dissipation in collisions. In addition, the mass and heat fluxes
vanish by symmetry reasons and the (uniform) pressure tensor is
the only nonzero flux of the problem. The relevant balance
equation is that for temperature, Eq.\ (\ref{2.11}), which reduces
to
\begin{equation}
\label{5.2}
\partial_tT+\frac{2}{dn}aP_{xy}=-(\zeta-\xi)T,
\end{equation}
where
\begin{equation}
\label{5.2.1} P_{xy}=\sum_{i=1}^2m_i\int\; \dd {\bf V} V_xV_y
f_i({\bf V})
\end{equation}
is the $xy$-element of the pressure tensor.

For a granular fluid under USF and in the absence of a
thermostatting force ($\xi=0$), the energy balance equation
(\ref{5.2}) leads to a steady state when the viscous heating
effect is exactly balanced by the collisional cooling. This
situation will be analyzed in Sec. \ref{sec10}. However, if for
instance the mixture is heated by the Gaussian thermostat
(\ref{3.5}) (with ${\bf v}\to {\bf V}$), then the viscous heating
still prevails so that the temperature increases in time. In this
case, the collision frequency $\nu_0(t)\propto \sqrt{T(t)}$ also
grows with $t$ and hence the reduced shear rate
$a^*(t)=a/\nu_0(t)$ (which is the relevant nonequilibrium
parameter of the problem) monotonically decreases in time. Under
these conditions, the system asymptotically achieves a regime
described by linear hydrodynamics and the (reduced) shear
viscosity $\eta^*=[\nu_0(t)/nT(t)]\eta(t)$ can be measured as
\begin{equation}
\label{5.3} \eta^*=-\lim_{t\to \infty}\frac{P_{xy}^*}{a^*},
\end{equation}
where $P_{xy}^*=P_{xy}/nT$. This procedure allows one to identify
the shear viscosity of a granular mixture {\em excited} by the
Gaussian external force (\ref{3.5}) and compare it with the
predictions given by the Chapman--Enskog method.

\begin{figure}
\centering
\includegraphics[width=0.7 \columnwidth]{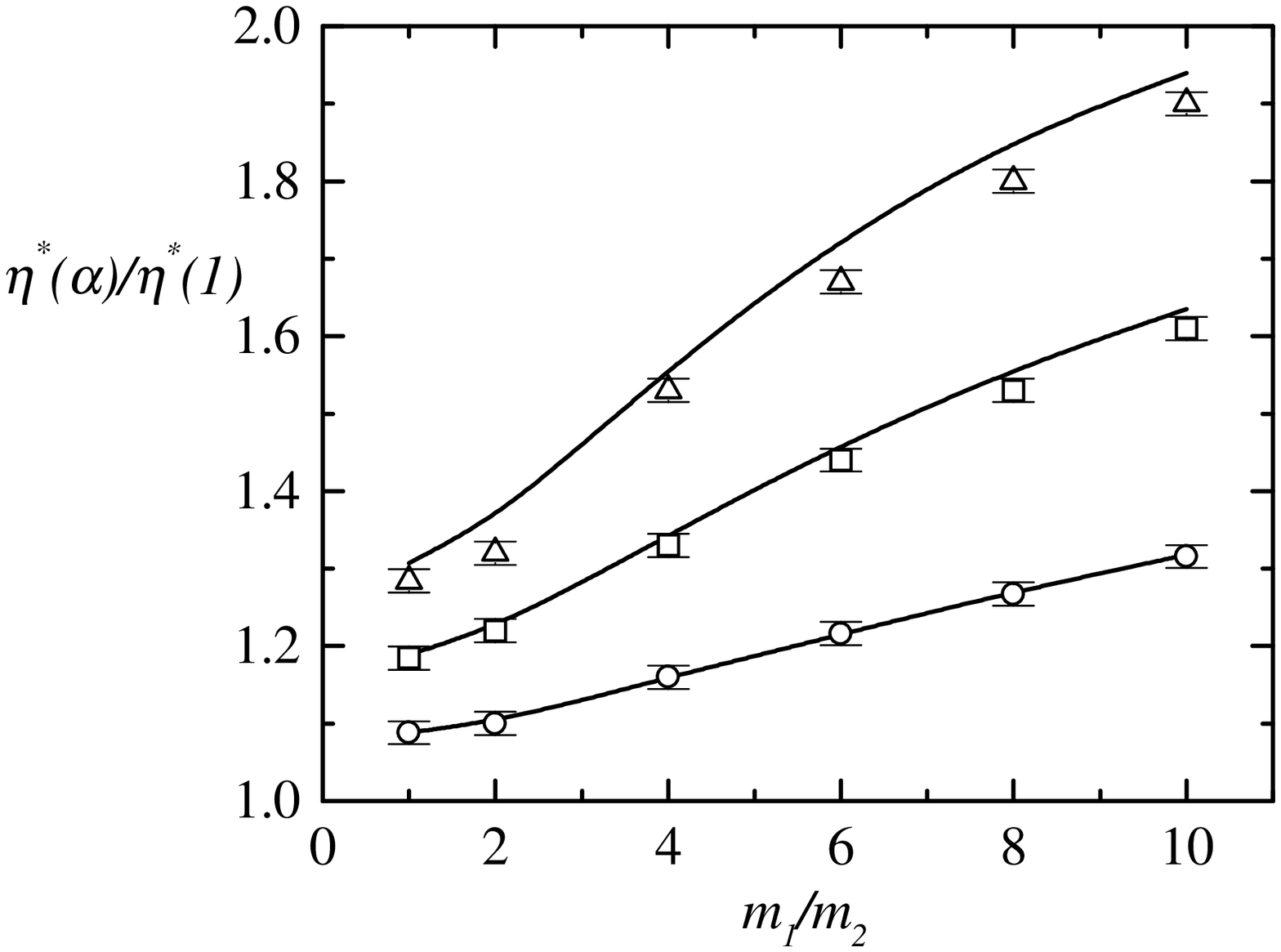}
\caption{Plot of the ratio $\eta^*(\alpha)/\eta^*(1)$ as a
function of the mass ratio $m_1/m_2$ for
$\sigma_1/\sigma_2=n_1/n_2=1$ and three different values of the
(common) coefficient of restitution $\alpha$: $\alpha=0.9$
(circles), $\alpha=0.8$ (squares), and $\alpha=0.7$ (triangles).
The lines are the theoretical predictions and the symbols refer to
the results obtained from the DSMC method. \label{fig3}}
\end{figure}
\begin{figure}
\centering
\includegraphics[width=0.7 \columnwidth]{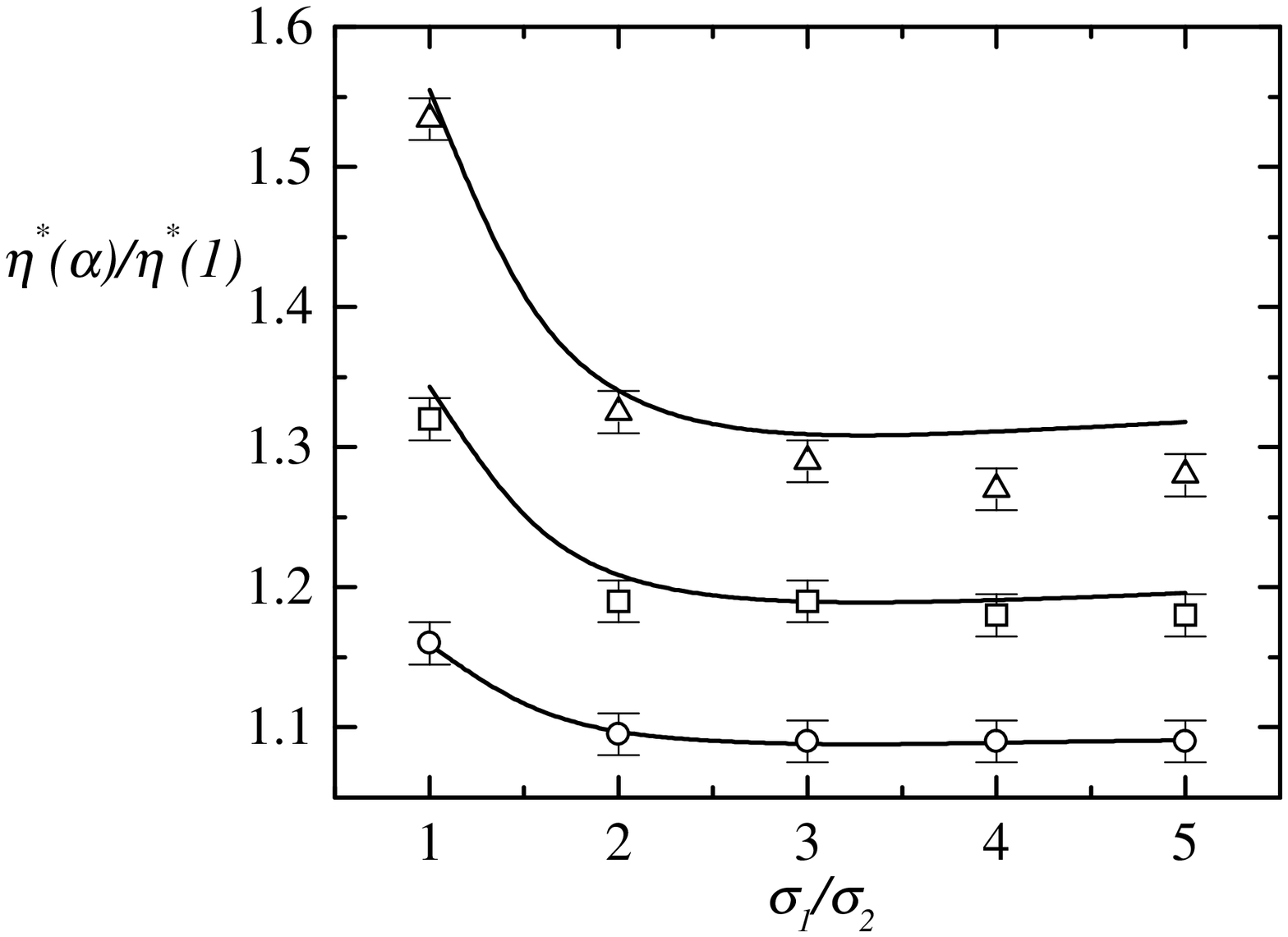}
\caption{Plot of the ratio $\eta^*(\alpha)/\eta^*(1)$ as a
function of the size ratio $\sigma_1/\sigma_2$ for $m_1/m_2=4$,
$n_1/n_2=1$ and three different values of the (common) coefficient
of restitution $\alpha$:  $\alpha=0.9$ (circles),  $\alpha=0.8$
(squares), and $\alpha=0.7$ (triangles). The lines are the
theoretical predictions and the symbols refer to the results
obtained from the DSMC method. \label{fig4}}
\end{figure}

In Fig.\ \ref{fig3}, we plot the ratio $\eta^*(\alpha)/\eta^*(1)$
versus the mass ratio $m_1/m_2$ in the case of hard spheres
($d=3$) for $\sigma_1/\sigma_2=1$, $x_1=\frac{1}{2}$, and three
different values of the (common) coefficient of restitution
$\alpha_{ij}\equiv \alpha$. Here, $\eta^*(1)$ refers to the
elastic value of the shear viscosity coefficient. Again, the
symbols represent the simulation data obtained by numerically
solving the Boltzmann equation \cite{MG03}, while the lines refer
to the theoretical results obtained from the Boltzmann equation in
the first Sonine approximation.  We see that in general the
deviation of $\eta^*(\alpha)$ from its functional form for elastic
collisions is quite important. This tendency becomes more
significant as the mass disparity increases. The agreement between
the first Sonine approximation and simulation is seen to be in
general excellent. This agreement is similar to the one previously
found in the monocomponent case \cite{BRC99,MSG05,GM02}. At a
quantitative level, the discrepancies between theory and
simulation tend to increase as the coefficient of restitution
decreases, although these differences are quite small (say, for
instance, around 2\% at $\alpha=0.7$ in the disparate mass case
$m_1/m_2=10$). The influence of the size ratio on the shear
viscosity is shown in Fig.\ \ref{fig4} for $m_1/m_2=4$ and
$x_1=\frac{1}{2}$ \cite{MG03}. We observe again a strong
dependence of the shear viscosity on dissipation. However, for a
given value of $\alpha$, the influence of $\sigma_1/\sigma_2$ on
$\eta^*$ is weaker than the one found before in Fig.\ \ref{fig3}
for the mass ratio. The agreement for both $\alpha=0.9$ and
$\alpha=0.8$ is quite good, except for the largest size ratio at
$\alpha=0.8$. These discrepancies become more significant as the
dissipation increases, especially for mixtures of particles of
very different sizes. In summary, according to the comparison
carried out in Figs.\ \ref{fig3} and \ref{fig4}, one can conclude
that the agreement between theory and simulation extends over a
wide range values of the coefficient of restitution, indicating
the reliability of the first Sonine approximation for describing
granular flows beyond the quasielastic limit.

\section{Einstein Relation in Granular Gases}
\label{sec6}

The results presented in Section \ref{sec5} give some support to
the validity of the hydrodynamic description to granular fluids.
However, in spite of this support some care is warranted in
extending properties of normal fluids to those with inelastic
collisions. Thus, for elastic collisions, in the case of an {\em
impurity} (tracer) particle immersed in a gas the response to an
external force on the impurity particle leads to a mobility
coefficient proportional to the diffusion coefficient. This is the
usual Einstein relation \cite{M89}, which is a consequence of the
fluctuation-dissipation theorem. A natural question is whether the
Einstein relation also applies for granular fluids.

To analyze it, let us consider the tracer limit ($x_1\to 0$) and
assume that the current of impurities ${\bf j}_1^{(1)}$ is only
generated by the presence of a weak concentration gradient $\nabla
x_1$ and/or a weak external field ${\bf F}_1$ acting only on the
impurity particles. Under these conditions, Eq.\ (\ref{4.3})
becomes
\begin{equation}
\label{6.1} {\bf j}_1^{(1)}=-m_1D\nabla x_1+\chi_{11} {\bf F}_1.
\end{equation}
The Einstein ratio $\epsilon'$ between the diffusion coefficient
$D$ and the mobility coefficient $\chi_{11}$ is defined as
\begin{equation}
\label{6.2} \epsilon'=m_1x_1\frac{D}{T\chi_{11}},
\end{equation}
where $T\simeq T_2$ in the tracer limit. For elastic collisions,
the Chapman--Enskog results yield $\epsilon'=1$. However, at
finite inelasticity the relationship between $D$ and $\chi_{11}$
is no longer simple and, as expected, the Chapman--Enskog
expressions for $D$ and $\chi_{11}$ in the case of an {\em
unforced} granular gas \cite{DG01} clearly show that $\epsilon'
\neq 1$. This means that the Einstein relation does not apply in
granular gases. The deviations of the (standard) Einstein ratio
$\epsilon'$ from unity has three distinct origins: the absence of
the Gibbs state (non-Gaussianity of the distribution function of
the HCS), time evolution of the granular temperature, and the
occurrence of different kinetic temperatures between the impurity
and gas particles. The second source of discrepancy can be avoided
if the system is driven by an external energy input to achieve a
stationary state. With respect to the third reason of violation,
this could also be partially eliminated if the temperature of the
gas $T$ is replaced by the temperature of the impurity $T_1$ in
the usual Einstein relation (\ref{6.2}). This change yields the
{\em modified} Einstein ratio
\begin{equation}
\label{6.3} \epsilon=m_1x_1\frac{D}{T_1\chi_{11}}.
\end{equation}
As a consequence, the only reason for which $\epsilon \neq 1$ is
due to the non-Maxwellian behavior of the HCS distribution. Given
that the deviations of the gas distribution $f_2^{(0)}$ from its
Maxwellian form are small \cite{GD99}, the discrepancies of
$\epsilon$ from unity could be difficult to detect in computer
simulations. This conclusion agrees with recent MD simulations
\cite{BLP04} of granular mixtures subjected to the stochastic
driving of the form (\ref{3.6}), where no deviations from the
(modified) Einstein relation $\epsilon=1$ have been observed for a
wide range of values of the coefficients of restitution and
parameters of the system.
\begin{figure}
\centering
\includegraphics[width=0.7 \columnwidth,angle=0]{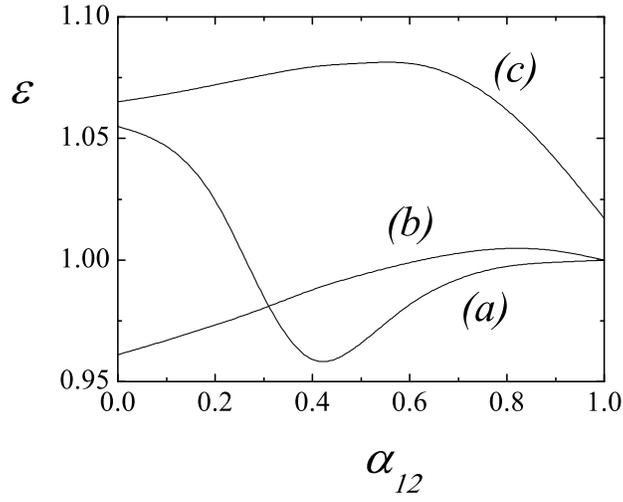}
\caption{Plot of the modified Einstein ratio $\epsilon$ versus the
coefficient of restitution $\alpha_{12}$ for the Gaussian
thermostat in the cases: (a) $\alpha_{22}=\alpha_{12}$,
$m_1/m_2=5$ and $\sigma_1/\sigma_2=1$; (b)
$\alpha_{22}=\alpha_{12}$, $m_1/m_2=0.5$ and
$\sigma_1/\sigma_2=1$; and (c) $\alpha_{22}=0.5$, $m_1/m_2=10$ and
$\sigma_1/\sigma_2=1$.} \label{fig5}
\end{figure}
\begin{figure}
\centering
\includegraphics[width=0.7 \columnwidth,angle=0]{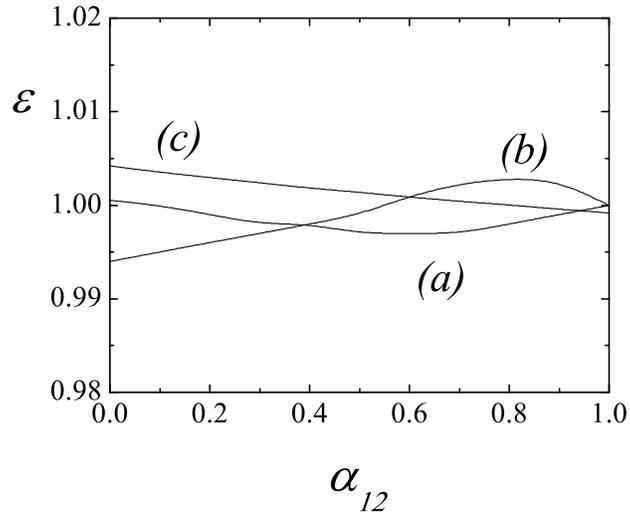}
\caption{Plot of the Einstein ratio $\epsilon$ versus the
coefficient of restitution $\alpha_{12}$ for the stochastic
thermostat in the cases: (a) $\alpha_{22}=\alpha_{12}$,
$m_1/m_2=5$ and $\sigma_1/\sigma_2=1$; (b)
$\alpha_{22}=\alpha_{12}$, $m_1/m_2=0.5$ and
$\sigma_1/\sigma_2=1$; and (c) $\alpha_{22}=0.5$, $m_1/m_2=10$ and
$\sigma_1/\sigma_2=1$.} \label{fig6}
\end{figure}

To illustrate the influence of dissipation on the Einstein ratio
 more generally, in Figs.\ \ref{fig5} and \ref{fig6}
the Einstein ratio as given by (\ref{6.3}) is plotted versus the
coefficient of restitution $\alpha_{12}$ for $\sigma_1/\sigma_2=1$
and different values of the mass ratio $m_1/m_2$ and the
coefficient of restitution $\alpha_{22}$. The results obtained by
using the Gaussian thermostat (\ref{3.5}) are shown in Fig.\
\ref{fig5}, while Fig.\ \ref{fig6} corresponds to the results
derived when the system is heated by the stochastic thermostat
(\ref{3.6}) \cite{G04}. We observe that in general $\epsilon \neq
1$, although its value is very close to unity, especially in the
case of the stochastic thermostat, where the deviations from the
Einstein relation are smaller than $1\%$. However, in the case of
the Gaussian thermostat the deviations from unity are about $8\%$,
which could be detected in computer simulations. Figures
\ref{fig5} and \ref{fig6} also show the fact that the transport
properties are affected by the thermostat introduced so that the
latter does not play a neutral role in the problem \cite{GM02}.

\section{Onsager's Reciprocal Relations in Granular Gases}
\label{sec7}

In the usual language of the linear irreversible thermodynamics
for {\em ordinary} fluids \cite{GM84}, the constitutive equations
(\ref{4.3}) and (\ref{4.5}) for the mass flux and heat flux  in
the absence of external forces can be written as
\begin{equation}
\mathbf{j}_{i}^{(1)}=-\sum_{j}L_{ij}\left( \frac{\nabla \mu
_{j}}{T}\right) _{T}-L_{iq}\frac{\nabla T}{T^{2}}-C_{p}\nabla p,
\label{7.1}
\end{equation}
\begin{equation}
\mathbf{J}_{q}^{(1)}=-L_{qq}\nabla T-\sum_{i}L_{qi}\left(
\frac{\nabla \mu _{i}}{T} \right) _{T}-C_{p}^{\prime }\nabla p,
\label{7.2}
\end{equation}
where
\begin{equation}
\label{7.2.1} {\bf J}_q^{(1)}\equiv {\bf
q}^{(1)}-\frac{d+2}{2}T\sum_i\frac{{\bf j}_i^{(1)}}{m_i}
\end{equation}
and
\begin{equation}
\left( \frac{\nabla \mu _{i}}{T}\right) _{T}=\frac{1}{m_{i}}\nabla
\ln (x_{i}p),  \label{7.3}
\end{equation}
$\mu _{i}$ being the chemical potential per unit mass. In Eqs.\
(\ref{7.1}) and (\ref{7.2}), the coefficients $L_{ij}$ are the
so-called Onsager phenomenological coefficients and the
coefficients $C_p$ and $C_p'$ can be expressed in terms of the
transport coefficients associated with the heat and mass fluxes.
For {\em elastic} fluids, Onsager showed \cite{GM84} that time
reversal invariance of the underlying microscopic equations of
motion implies important constraints on the above set of transport
coefficients, namely
\begin{equation}
L_{ij}=L_{ji},\quad L_{iq}=L_{qi},\quad C_{p}=C_{p}^{\prime }=0.
\label{7.4}
\end{equation}
The first two symmetries are called reciprocal relations as they
relate transport coefficients for different processes. The last
two are statements that the pressure gradient does not appear in
any of the fluxes, even though it is admitted by symmetry. Even
for a one component fluid, Onsager's theorem is significant as it
leads to a {\em new} contribution to the heat flux proportional to
the density gradient \cite{BDKS98}. Since there is no time
reversal symmetry for granular fluids, Eqs. (\ref{7.4}) cannot be
expected to apply. However, since explicit expressions for all
transport coefficients are at hand, the quantitative extent of the
violation can be explored.

To make connection with the expressions (\ref{4.3}) and
(\ref{4.5}) for the mass and heat fluxes, respectively, it is
first necessary to transform Eqs.\ (\ref{7.1})--(\ref{7.2.1}) to
the variables $x_{1},$ $p,$ $T.$ Since $\nabla x_{1}=-\nabla
x_{2}$, Eq.\ (\ref{7.3}) implies
\begin{equation}
\frac{(\nabla \mu _{1})_{T}-(\nabla \mu _{2})_{T}}{T}=\frac{n\rho
}{\rho _{1}\rho _{2}}\left[ \nabla x_{1}+\frac{n_{1}n_{2}}{n\rho }
(m_{2}-m_{1})\nabla \ln p\right] .  \label{7.5}
\end{equation}
The coefficients $\{L_{ij}, L_{iq}, L_{qi}, L_{qq}, C_p, C_p'\}$
then can be easily obtained in terms of the Navier-Stokes
transport coefficients introduced in Sec.\ \ref{sec4}. The result
is \cite{GMD06}
\begin{equation}
L_{11}=-L_{12}=-L_{21}=\frac{m_{1}m_{2}\rho _{1}\rho _{2}}{\rho
^{2}}D,\quad L_{1q}=\rho TD^{\prime },  \label{7.6}
\end{equation}
\begin{equation}
\label{7.7} L_{q1}=-L_{q2}=\frac{T^{2}\rho _{1}\rho _{2}}{n\rho
}D^{\prime \prime }- \frac{d+2}{2}\frac{T\rho _{1}\rho _{2}}{\rho
^{2}}(m_{2}-m_{1})D,
\end{equation}
\begin{equation}
\label{7.7.1} L_{qq}=\lambda -\frac{d+2}{2}\rho
\frac{m_{2}-m_{1}}{m_{1}m_{2}}D^{\prime },
\end{equation}
\begin{equation}
C_{p}\equiv \frac{\rho}{p} D_{p}-\frac{\rho _{1}\rho _{2}}{p\rho ^{2}}%
(m_{2}-m_{1})D,  \label{7.8}
\end{equation}
\begin{equation}
C_{p}^{\prime}\equiv L-\frac{d+2}{2}\frac{T}{p}\frac{m_{2}-m_{1}}{m_{1}m_{2}}%
C_{p}- \frac{n_{1}n_{2}}{np\rho }T^{2}(m_2-m_1)D^{\prime \prime}.
\label{7.9}
\end{equation}

Onsager's relation $L_{12}=L_{21}$ holds since the diffusion
coefficient $D$ is symmetric under the change $1\leftrightarrow 2$
\cite{GD02}. However, in general $L_{1q}\neq L_{q1}$, $C_p \neq
0$, and $C_p'\neq 0$ \cite{GMD06}. The Chapman-Enskog results
\cite{GD02} show that there are only two limit cases for which
$L_{1q}-L_{q1}=C_p=C_p'=0$: (i) the elastic limit
($\alpha_{ij}=1$) with arbitrary values of masses, sizes and
composition and (ii) the case of mechanically equivalent particles
with arbitrary values of the (common) coefficient of restitution
$\alpha\equiv \alpha_{ij}$. Beyond these limit cases, Onsager's
relations do not apply. At macroscopic level the origin of this
failure is due to the cooling of the reference state as well as
the occurrence of different kinetic temperatures for both species.

\section{Linearized Hydrodynamic Equations and Stability
of the Homogeneous Cooling State} \label{sec8}

As shown in Sec.\ \ref{sec4}, the Navier-Stokes constitutive
equations (\ref{4.3})--(\ref{4.5}) have been expressed in terms of
a set of experimentally accessible fields such as the composition
of species $1$, $x_1$, the pressure $p$, the mean flow field ${\bf
u}$, and the granular temperature $T$. In terms of these variables
and in the absence of external forces, the macroscopic balance
equations (\ref{2.9})--(\ref{2.11}) become
\begin{equation}
D_{t}x_{1}+\frac{\rho }{n^{2}m_{1}m_{2}}\nabla \cdot
\mathbf{j}_{1}=0\;, \label{8.1}
\end{equation}
\begin{equation}
D_{t}p+p\nabla \cdot \mathbf{u}+\frac{2}{d}\left( \nabla \cdot
\mathbf{q}+ \mathsf{P}:\nabla \mathbf{u}\right) =-\zeta p,
\label{8.2}
\end{equation}
\begin{equation}
D_{t}\mathbf{u}+\rho ^{-1}\nabla \cdot \mathsf{P}=0\;, \label{8.3}
\end{equation}
\begin{equation}
D_{t}T-\frac{T}{n}\sum_{i}\frac{\nabla \cdot
\mathbf{j}_{i}}{m_{i}}+\frac{2}{dn}\left( \nabla \cdot
\mathbf{q}+\mathsf{P}:\nabla \mathbf{u}\right) =-\zeta T\;.
\label{8.4}
\end{equation}
When the expressions (\ref{4.3})--(\ref{4.5}) for the fluxes and
the cooling rate $\zeta\to \zeta^{(0)}$ are substituted into the
above exact balance equations (\ref{8.1})--(\ref{8.4}) one gets a
closed set of hydrodynamic equations for $x_1, {\bf u}, T,$ and
$p$. These are the Navier--Stokes hydrodynamic equations for a
binary granular mixture:
\begin{equation}
D_{t}x_{1}=\frac{\rho }{n^{2}m_{1}m_{2}}\nabla \cdot \left(
\frac{m_{1}m_{2}n }{\rho }D\nabla x_{1}+\frac{\rho }{p}D_{p}\nabla
p+\frac{\rho }{T}D' \nabla T\right) \;,  \label{8.5}
\end{equation}
\begin{eqnarray}
\left( D_{t}+\zeta \right) p+\frac{d+2}{d}p\nabla \cdot {\bf u}
&=&\frac{2}{d}\nabla \cdot \left( T^{2}D^{\prime \prime}\nabla
x_{1}+L\nabla p+\lambda
\nabla T\right)  \nonumber \\
&&+\frac{2}{d}\eta \left( \nabla _{\ell }u_{k}+\nabla _{k}u_{\ell
}-\frac{2}{ d}\delta _{k\ell }\nabla \cdot {\bf u}\right) \nabla
_{\ell }u_{k},\nonumber\\ \label{8.6}
\end{eqnarray}
\begin{eqnarray}
\left( D_{t}+\zeta \right) T+\frac{2}{d n}p\nabla \cdot {\bf u}
&=& -\frac{T}{n}\frac{m_2-m_1}{m_1m_2} \nabla \cdot \left(
\frac{m_{1}m_{2}n}{\rho }D\nabla x_{1}+\frac{\rho }{p}D_{p}\nabla
p  \right.\nonumber \\
&&\left.+ \frac{\rho}{T}D^{\prime }\nabla T\right)+\frac{2}{d
n}\nabla \cdot \left( T^{2}D^{\prime \prime}\nabla
x_{1}+L\nabla p+\lambda \nabla T\right)  \nonumber \\
&&+\frac{2}{d n}\eta \left( \nabla _{\ell }u_{k}+\nabla
_{k}u_{\ell }-\frac{2 }{d}\delta _{k\ell }\nabla \cdot
\mathbf{u}\right) \nabla _{\ell }u_{k}, \label{8.7}
\end{eqnarray}
\begin{equation}
D_{t}u_{\ell}+\rho^{-1}\nabla_{\ell}p=\rho ^{-1}\nabla _{k}\eta
\left( \nabla _{\ell }u_{k}+\nabla _{k}u_{\ell }-\frac{2}{d}\delta
_{k\ell }\nabla \cdot {\bf u}\right) \;. \label{8.8}
\end{equation}
For the chosen set of fields, $n=p/T$ and $\rho =p\left[ \left(
m_{1}-m_{2}\right) x_{1}+m_{2}\right] /T$. These equations are
exact to second order in the spatial gradients for a low density
Boltzmann gas. Note that in Eqs.\ (\ref{8.5})--(\ref{8.8}) the
second order contributions to the cooling rate have been
neglected. These second order terms have been calculated for a
monocomponent fluid \cite{BDKS98} and found to be very small
relative to corresponding terms from the fluxes. Consequently,
they have not been considered in the hydrodynamic equations
(\ref{8.5})--(\ref{8.8}).

One of the main peculiarities of the granular gases (in contrast
to ordinary fluids) is the existence of non-trivial solutions to
the Navier--Stokes equations (\ref{8.5})--(\ref{8.8}), even for
spatially homogeneous states,
\begin{equation}  \label{8.9}
\partial _{t}x_{1H}=0,\quad \partial _{t}{\bf u}_{H}={\bf 0},
\end{equation}
\begin{equation}
\left[ \partial _{t}+\zeta \left( x_{1H},T_{H},p_{H}\right)
\right] T_{H}=0, \hspace{0.3in}\left[ \partial _{t}+\zeta \left(
x_{1H},T_{H},p_{H}\right) \right] p_{H}=0,  \label{8.10}
\end{equation}
where the subscript $H$ denotes the homogeneous state. Since the
dependence of the cooling rate $\zeta \left(
x_{1H},T_{H},p_{H}\right)$ on $ x_{1H},T_{H},p_{H}$ is known
\cite{GD99,GM06}, these first order nonlinear equations can be
solved for the time dependence of the homogeneous state. The
result is the familiar Haff cooling law for $T(t)$ at constant
density \cite{H83,BP04}:
\begin{equation}
\label{8.11} T_H(t)=\frac{T_H(0)}{\left[1+\zeta(0)t/2\right]^2}.
\end{equation}
As said before, each partial temperature $T_i(t)$ has the same
time dependence but with a different value \cite{GD99},
\begin{equation}
\label{8.12} T_{1H}(t)=\frac{\gamma}{1+x_1(\gamma-1)}T_H(t),\quad
T_{2H}(t)=\frac{1}{1+x_1(\gamma-1)}T_H(t),
\end{equation}
where $\gamma=T_{1H}(t)/T_{2H}(t)$ is the time-independent
temperature ratio.

Nevertheless, the homogeneous cooling state (HCS) is unstable to
sufficiently long wavelength perturbations. For systems large
enough to support such spontaneous fluctuations, the HCS becomes
{\em inhomogeneous} at long times. This feature was first observed
in MD simulations of free monocomponent gases \cite{GZ93}. In MD
simulations the inhomogeneities may grow by the formation of
clusters, ultimately aggregating to a single large cluster
\cite{LH99}; if cluster growth is suppressed, a vortex field may
grow to the system size where periodic boundary conditions can
induce a transition to a state with macroscopic shear. The
mechanism responsible for the growth of inhomogeneities can be
understood at the level of the Navier--Stokes hydrodynamics, where
{\em linear} stability analysis shows two shear modes and a heat
mode to be unstable \cite{BDKS98,BRC99, BP04, G05}.

The objective here is to extend this analysis to the case of a
binary mixture. To do that, we perform a linear stability analysis
of the nonlinear hydrodynamic equations (\ref{8.5})--(\ref{8.8})
with respect to this HCS for small initial spatial perturbations.
For ordinary fluids such perturbations decay in time according to
the hydrodynamic modes of diffusion (shear, thermal, mass) and
damped sound propagation \cite{HM86,RL77,BY80}. For inelastic
collisions, the analysis is for fixed coefficients of restitution
in the long wavelength limit. As will be seen below, the
corresponding modes for a granular mixture are then quite
different from those for ordinary mixtures. In fact, an
alternative study with fixed long wavelength and coefficients of
restitution approaching unity yields the usual ordinary fluid
modes. Consequently, the nature of the hydrodynamic modes is non
uniform with respect to the inelasticity and the wavelength of the
perturbation.

Let us assume that the deviations $\delta y_{\alpha}({\bf
r},t)=y_{\alpha}({\bf r},t)-y_{H \alpha}(t)$ are small. Here,
$\delta y_{\alpha}({\bf r},t)$ denotes the deviation of $\{x_1,
{\bf u}, T, p\}$ from their values in the HCS. If the initial
spatial perturbation is sufficiently small, then for some initial
time interval these deviations will remain small and the
hydrodynamic equations (\ref{8.5})--(\ref{8.8}) can be linearized
with respect to $\delta y_{\alpha }(\mathbf{r},t)$. This leads to
a set of partial differential equations with coefficients that are
independent of space but which depend on time. As in the
monocomponent case \cite{BDKS98,G05}, this time dependence can be
eliminated through a change in the time and space variables, and a
scaling of the hydrodynamic fields. We introduce the following
dimensionless space and time variables:
\begin{equation}
\label{8.13} \tau=\int_{0}^{t} \dd t' \nu_{0H}(t'),\quad {\bf
s}=\frac{\nu_{0H}(t)}{v_{0H}(t)}{\bf r},
\end{equation}
where $\nu_{0H}(t)$ is an effective collision frequency for hard
spheres and $v_{0H}=\sqrt{2T_H(m_1+m_2)/m_1m_2}$. Since $\{x_{1H},
{\bf u}_H, T_H, p_H\}$ are evaluated in the HCS, then Eqs.\
(\ref{8.9}) and (\ref{8.10}) hold. A set of Fourier transformed
dimensionless variables are then introduced as
\begin{equation}
\label{8.15} \rho_{{\bf k}}(\tau)=\frac{\delta x_{1{\bf
k}}(\tau)}{x_{1H}}, \quad {\bf w}_{{\bf k}}(\tau)=\frac{\delta
{\bf u}_{{\bf k}}(\tau)}{v_{0H}(\tau)},\quad \theta_{{\bf
k}}(\tau)=\frac{\delta T_{{\bf k}}(\tau)}{T_{H}(\tau)}, \quad
\Pi_{{\bf k}}(\tau)=\frac{\delta p_{{\bf k}}(\tau)}{p_{H}(\tau)},
\end{equation}
where $\delta y_{\alpha{\bf k}}\equiv \{\delta x_{1{\bf k}},
\delta {\bf u}_{\bf k}, \delta T_{\bf k}, \delta p_{\bf k}\}$ is
defined as
\begin{equation}
\label{8.16} \delta y_{\alpha{\bf k}}(\tau)=\int \dd{\bf s}\;
e^{-i{\bf k}\cdot {\bf s}}\delta y_{\alpha}({\bf s},\tau).
\end{equation}
Note that here the wave vector ${\bf k}$ is dimensionless.

In terms of the above variables, the transverse velocity
components ${\bf w}_{\mathbf{k}\perp}={\bf w}_{\bf k}-({\bf
w}_{\bf k}\cdot \widehat{{\bf k}})\widehat{{\bf k}}$ (orthogonal
to the wave vector ${\bf k}$) decouple from the other four modes
and hence can be obtained more easily. They obey the equation
\begin{equation}
\left( \frac{\partial }{\partial \tau }-\frac{\zeta ^{*}}{2}+\eta
^{* }k^{2}\right) {\bf w}_{{\bf k}\perp}={\bf 0},  \label{8.17}
\end{equation}
where $\zeta ^{*}=\zeta _{H}/\nu_{0H}$ and
\begin{equation}
\label{8.18} \eta^*=\frac{\nu_{0H}}{\rho_H v_{0H}^2}\eta,
\end{equation}
where $\rho_H=m_1n_{1H}+m_2n_{2H}$. The solution for ${\bf
w}_{{\bf k}\perp}(\tau)$ reads
\begin{equation}
\label{8.19} {\bf w}_{{\bf k}\perp}(\tau)={\bf w}_{{\bf
k}\perp}(0)\exp[s_{\perp}(k)\tau],
\end{equation}
where
\begin{equation}
\label{8.20} s_{\perp}(k)=\frac{1}{2}\zeta^*-\eta^* k^2.
\end{equation}
This identifies $d-1$ shear (transversal) modes. We see from Eq.\
(\ref{8.20}) that there exists a critical wave number
$k_{\perp}^{\text{c}}$ given by
\begin{equation}
\label{8.21}
k_{\perp}^{\text{c}}=\left(\frac{\zeta^*}{2\eta^*}\right)^{1/2}.
\end{equation}
This critical value separates two regimes: shear modes with $k\geq
k_{\perp}^{\text{c}}$ always decay while those with $k<
k_{\perp}^{\text{c}}$ grow exponentially.

The remaining modes are called longitudinal modes. They correspond
to the set $\{ \rho_{{\bf k}}, \theta_{{\bf k}}, \Pi_{{\bf
k}},w_{{\bf k}||} \}$ where the longitudinal velocity component
(parallel to ${\bf k}$) is $w_{{\bf k}||}={\bf w}_{\bf k}\cdot
\widehat{{\bf k}}$. These modes are the solutions of the linear
equation \cite{GMD06}
\begin{equation}
\frac{\partial \delta z_{\alpha {\bf k}}(\tau )}{\partial \tau
}=\left( M_{\alpha \beta }^{(0)}+ikM_{\alpha \beta
}^{(1)}+k^{2}M_{\alpha \beta }^{(2)}\right) \delta z_{\beta{\bf k}
}(\tau ), \label{8.22}
\end{equation}
where $\delta z_{\alpha {\bf k}}(\tau )$ denotes now the four
variables $\left\{ \rho _{{\bf k}},\theta _{{\bf k}},\Pi _{{\bf
k}},w_{{\bf k}||}\right\}$. The matrices in Eq.\ (\ref{8.22}) are
given by
\begin{equation}
{\sf M}^{(0)}=\left(
\begin{array}{cccc}
0 & 0 & 0 & 0 \\
-x_{1}\left(\frac{\partial \zeta ^*}{\partial x_{1}}\right) _{T,p}
&
\frac{1}{2}\zeta ^* & -\zeta ^{*} & 0 \\
-x_{1}\left(\frac{\partial \zeta ^*}{\partial x_{1}}\right) _{T,p}
&
\frac{1}{2}\zeta ^{*} & -\zeta ^{*} & 0 \\
0 & 0 & 0 & \frac{1}{2}\zeta ^{*}
\end{array}
\right),   \label{8.23}
\end{equation}
\begin{equation}
{\sf M}^{(1)}=\left(
\begin{array}{cccc}
0 & 0 & 0 & 0 \\
0 & 0 & 0 & -\frac{2}{d} \\
0 & 0 & 0 & -\frac{d+2}{d}\\
0 & 0 & -\frac{1}{2}\frac{\mu_{12}}{x_{1}\mu +x_{2}} & 0
\end{array}
\right),   \label{8.24}
\end{equation}
\begin{equation}
{\sf M}^{(2)}=\left(
\begin{array}{cccc}
-D^{*} & -x_{1}^{-1}D'^{*} & -x_{1}^{-1}D_{p}^{*} & 0\\
M_{21}^{(2)}  & M_{22}^{(2)} &
M_{23}^{(2)} & 0 \\
-\frac{2}{d}x_{1}D''^{*} & -\frac{2}{d}\lambda ^{*} &
-\frac{2}{d}L^{*} & 0 \\
0 & 0 & 0 & -\frac{2}{d}(d-1)\eta ^{*}
\end{array}
\right),   \label{8.25}
\end{equation}
where
\begin{equation}
\label{8.25.1} M_{21}^{(2)}=-x_{1}\left(
\frac{2}{d}D''^{*}-\frac{1-\mu }{x_{1}\mu+x_{2}}D^{*}\right),
\end{equation}
\begin{equation}
\label{8.25.2}
M_{22}^{(2)}=\frac{1-\mu}{x_{1}\mu+x_{2}}D'^{*}-\frac{
2}{d}\lambda ^{*},
\end{equation}
\begin{equation}
\label{8.25.3}
M_{23}^{(2)}=-\frac{2}{d}L^{*}+\frac{1-\mu}{x_{1}\mu+x_{2}}
D_{p}^{*}.
\end{equation}
In these equations, $\mu=m_1/m_2$, $x_i=n_{iH}/n_H$,  and we have
introduced the reduced Navier--Stokes transport
coefficients\footnote{Note that the definition for the reduced
diffusion coefficient $D^*$ given here differs from the one
introduced in Sec.\ \ref{sec5.1}.}
\begin{equation}
\label{8.26} D^*=\frac{\nu_{0H}}{n_Hv_{0H}^2}D,\quad
D_p^*=\frac{\rho_H^2\nu_{0H} }{m_1m_2n_H^2v_{0H}^2}D_p,\quad
D'^{*}=\frac{\rho_H^2\nu_{0H} }{m_1m_2n_H^2v_{0H}^2}D',
\end{equation}
\begin{equation}
\label{8.27} D''^{*}=\frac{\nu_{0H} T_H}{n_Hv_{0H}^2}D'',\quad
L^*=\frac{\nu_{0H} }{v_{0H}^2}L,\quad \lambda^{*}=\frac{\nu_{0H}
}{n_Hv_{0H}^2}\lambda.
\end{equation}
\begin{figure}
\centering
\includegraphics[width=0.7 \columnwidth,angle=0]{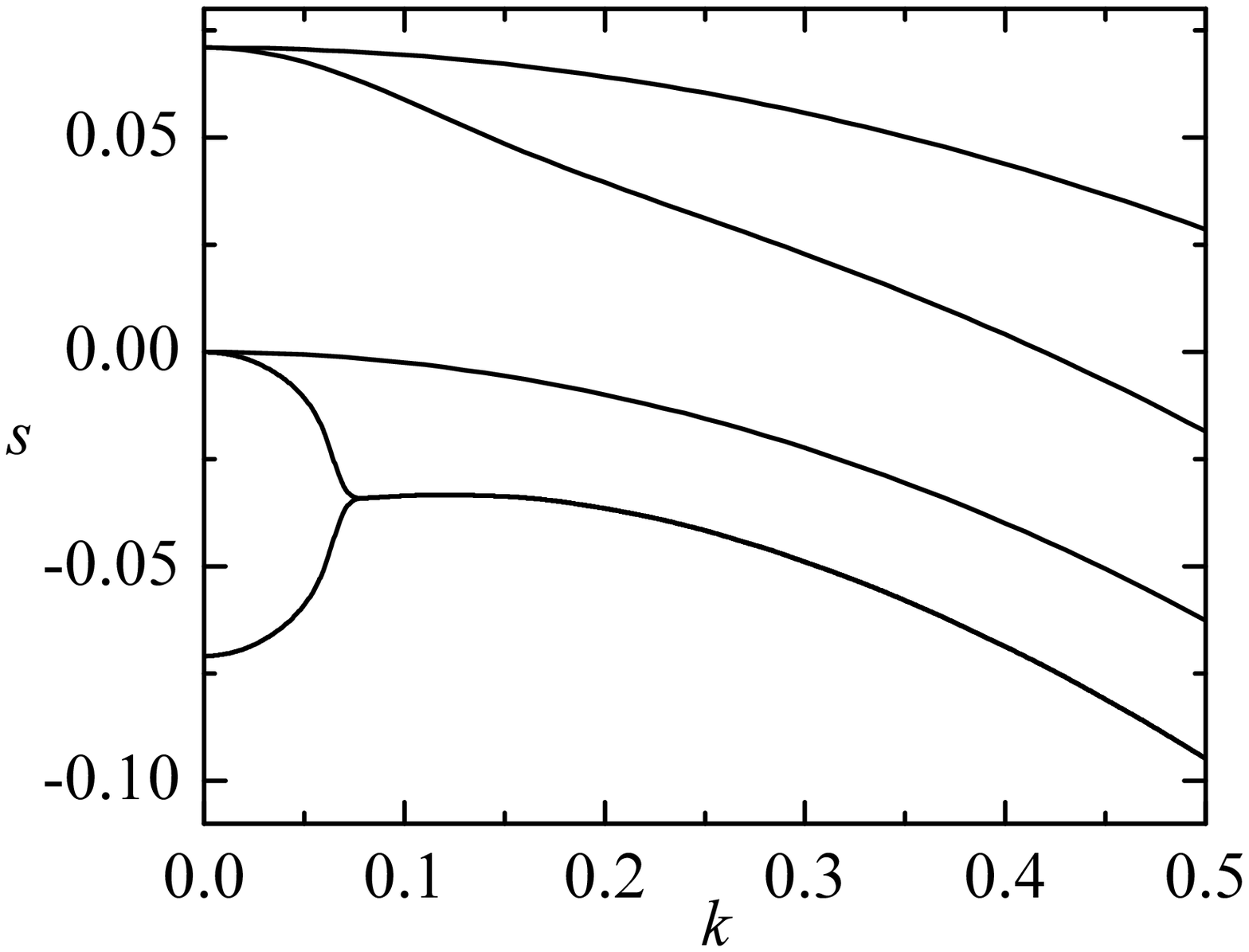}
\caption{Dispersion relations for $\protect\alpha=0.9$, $x_{1}=0.2$, $%
\protect\omega =1$ and $\protect\mu =4$.} \label{fig7}
\end{figure}

The longitudinal modes have the form $\exp[s_{n}(k)\tau]$ with
$n=1,2,3,4$, where $s_n(k)$ are the eigenvalues of the matrix
${\sf M}(k)={\sf M}^{(0)}+i k {\sf M}^{(1)}+k^2 {\sf M}^{(2)}$,
namely, they are the solutions of the quartic equation
\begin{equation}
\label{8.28} \det |{\sf M}-s\openone|=0.
\end{equation}
The solution to (\ref{8.28}) for arbitrary values of $k$ is quite
intricate. It is instructive to consider first the solutions to
these equations in the extreme long wavelength limit, $k=0$. In
this case, they are found to be the eigenvalues of the matrix of
${\sf M}^{(0)}$:
\begin{equation}
s_{n}^{(0)}=\left( 0,0,-\frac{1 }{2} \zeta ^{\ast
},\frac{1}{2}\zeta ^{\ast }\right). \label{8.29}
\end{equation}
Hence, at asymptotically long wavelengths (${\bf k}=0$) the
spectrum of the linearized hydrodynamic equations (both transverse
and longitudinal) is comprised of a decaying mode at $-\zeta
^{*}/2$, a two-fold degenerate mode at $0$, and a $d$-fold
degenerate unstable mode at $\zeta ^{*}/2$. Consequently, some of
the solutions are unstable. The two zero eigenvalues represent
marginal stability solutions, while the negative eigenvalue gives
stable solutions. For general initial perturbations all modes are
excited. These modes correspond to evolution of the fluid due to
uniform perturbations of the HCS, i.e., a global change in the HCS
parameters. The unstable modes are seen to arise from the initial
perturbations $w_{\mathbf{k}\perp }(0)$ or $ w_{\mathbf{k}||}(0)$.
The marginal modes correspond to changes in the composition at
fixed pressure, density, and velocity, and to changes in $
\Pi_{\mathbf{k}}-\theta_{\mathbf{k}}$ at constant composition and
velocity. The decaying mode corresponds to changes in the
temperature or pressure for $
\Pi_{\mathbf{k}}=\theta_{\mathbf{k}}$. The unstable modes may
appear trivial since they are due entirely to the normalization of
the fluid velocity by the time dependent thermal velocity.
However, this normalization is required by the scaling of the
entire set of equations to obtain time independent coefficients.

The real parts of the modes $s_{\perp}(k)$ and $s_n(k)$ is
illustrated in Fig.\ \ref{fig7} in the case of hard spheres
($d=3$) for $\alpha\equiv \alpha_{ij}=0.9$, $\sigma_1/\sigma_2=1$,
$x_1=0.2$, and $m_1/m_2=4$. The $k=0$ values correspond to five
hydrodynamic modes with two different degeneracies. The shear mode
degeneracy remains at finite $k$ but the other is removed at any
finite $k$. At sufficiently large $k$ a pair of real modes become
equal and become a complex conjugate pair at all larger wave
vectors, like a sound mode. The smallest of the unstable modes is
that associated with the longitudinal velocity, which couples to
the scalar hydrodynamic fields. It becomes negative at a wave
vector smaller than that of Eq.\ (\ref{8.21}) and gives the
threshold for development of spatial instabilities.

The results obtained here for mixtures show no new surprises
relative to the case for a monocomponent gas
\cite{BDKS98,BP04,G05}, with only the addition of the stable mass
diffusion mode. Of course, the quantitative features can be quite
different since there are additional degrees of freedom with the
parameter set $\left\{ x_{1H},m_{1}/m_{2},\sigma _{1}/\sigma
_{2},\alpha _{ij}\right\}$. Also, the manner in which these linear
instabilities are enhanced by the nonlinearities may be different
from that for the one component case \cite{BRC99bis}.

\section{Segregation in Granular Binary Mixtures: Thermal
Diffusion} \label{sec9}

The analysis of the linearized hydrodynamic equations for a
granular binary mixture has shown that the resulting equations
exhibit a long wavelength instability for $d$ of the modes. These
instabilities lead to the spontaneous formation of velocity
vortices and density clusters when the system evolves freely. A
phenomenon related with the density clustering is the separation
or species segregation. Segregation and mixing of dissimilar
grains is perhaps one of the most interesting problems in agitated
granular mixtures. In some processes it is a desired and useful
effect to separate particles of different types, while in other
situations it is undesired and can be difficult to control. A
variety of mechanisms have been proposed to describe the
separation of particles of two sizes in a mixture of vertically
shaken particles. Different mechanisms include void filling,
static compressive force, convection, condensation, thermal
diffusion, interstitial gas forcing, friction, and buoyancy
\cite{K04}. However, in spite of the extensive literature
published in the past few years on this subject, the problem is
not completely understood yet. Among the different competing
mechanisms, thermal diffusion becomes one of the most relevant at
large shaking amplitude where the sample of macroscopic grains
resembles a granular gas. In this regime, binary collisions
prevail and kinetic theory can be quite useful to analyze the
physical mechanisms involved in segregation processes.

Thermal diffusion is caused by the relative motion of the
components of a mixture because of the presence of a temperature
gradient. Due to this motion, concentration gradients subsequently
appear in the mixture producing diffusion that tends to oppose
those gradients. A steady state is finally achieved in which the
separation effect arising from thermal diffusion is compensated by
the diffusion effect. In these conditions, the so-called thermal
diffusion factor $\Lambda_{ij}$ characterizes the amount of
segregation parallel to the temperature gradient. In this Section,
the thermal diffusion factor is determined from the
Chapman--Enskog solution described before.

To make some contact with experiments, let us assume that the
binary granular mixture is in the presence of the gravitational
field ${\bf g}=-g \hat{{\bf e}}_z$, where $g$ is a positive
constant and $\hat{{\bf e}}_z$ is the unit vector in the positive
direction of the $z$ axis. In experiments \cite{SUKSS06}, the
energy is usually supplied by vibrating horizontal walls so that
the system reaches a steady state. Here, instead of considering
oscillating boundary conditions, particles are assumed to be
heated by the action of the stochastic driving force (\ref{3.6}),
which mimics a thermal bath. As said above, although the relation
between this driven idealized method with the use of locally
driven wall forces is not completely understood, it must be
remarked that in the case of boundary conditions corresponding to
a sawtooth vibration of one wall the condition to determine the
temperature ratio coincides with the one derived from the
stochastic force \cite{DHGD02}. The good agreement between theory
and simulation found in Fig.\ \ref{fig1} for the temperature ratio
confirms this expectation.

The thermal diffusion factor $\Lambda_{ij}$ ($i\neq j$) is defined
at the steady state in which the mass fluxes ${\bf j}_i$ vanish.
Under these conditions, the factor $\Lambda_{ij}$ is given through
the relation \cite{KCM83}
\begin{equation}
\label{9.1} -\Lambda_{ij}\nabla \ln T=\frac{1}{x_ix_j}\nabla x_i
,\quad \Lambda_{ij}+\Lambda_{ji}=0.
\end{equation}
The physical meaning of $\Lambda_{ij}$ can be described by
considering a granular binary mixture held between plates at
different temperatures $T$ (top plate) and $T'$ (bottom plate)
under gravity. For the sake of concreteness, we will assume that
gravity and thermal gradient point in parallel directions, i.e.,
the bottom is hotter than the top ($T'>T$). In addition,  without
loss of generality, we also assume that $\sigma_1>\sigma_2$.  In
the steady state, Eq.\ (\ref{9.1}) describes how the thermal field
is related to the composition of the mixture. Assuming that
$\Lambda_{12}$ is constant over the relevant ranges of temperature
and composition, integration of Eq.\ ({\ref{9.1}) yields
\begin{equation}
\label{9.2} \ln \frac{x_1x_2'}{x_2x_1'}=\Lambda_{12}\ln
\frac{T'}{T},
\end{equation}
where $x_i$ refers to the mole fraction of species $i$ at the top
plate and $x_i'$ refers to the mole fraction of species $i$ at the
bottom plate. Consequently, according to Eq.\ (\ref{9.2}), if
$\Lambda_{12}>0$, then $x_1'<x_1$, while if $\Lambda_{12}<0$, then
$x_1'>x_1$. In summary, when $\Lambda_{12}>0$, the larger
particles accumulate at the top of the sample (cold plate), while
if $\Lambda_{12}<0$, the larger particles accumulate at the bottom
of the sample (hot plate). The former situation is referred to as
the Brazil-nut effect (BNE) while the latter is called the reverse
Brazil-nut effect (RBNE).

The RBNE was first observed  by Hong {\em et al.} \cite{HQL01} in
MD simulations of vertically vibrated systems. They proposed a
very simple segregation criterion that was later confirmed by
Jenkins and Yoon \cite{JY02} by using kinetic theory. More
recently, Breu {\em et al. }\cite{BEKR03} have experimentally
investigated conditions under which the large particles sink to
the bottom and claim that their experiments confirm the theory of
Hong {\em et al.} \cite{HQL01} provided a number of conditions are
chosen carefully. In addition to the vertically vibrated systems,
some works have also focused in the last few years on horizontally
driven systems showing some similarities to the BNE and its
reverse form \cite{horizontal}. However, it is important to note
that the criterion given in Ref.\ \cite{HQL01} is based on some
drastic assumptions: elastic particles, homogeneous temperature,
and energy equipartition. These conditions preclude a comparison
of the kinetic theory derived here with the above simulations.
\begin{figure}
\centering
\includegraphics[width=0.7 \columnwidth,angle=0]{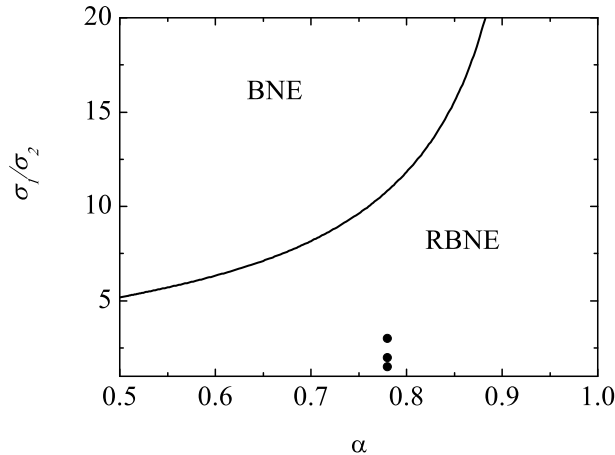}
\caption{Phase diagram for BNE/RBNE for mixtures constituted by
spheres ($d=3$) of the same mass density and equal total volumes
of large and small particles. The data points represent the MD
simulation results \cite{SUKSS06} for $\alpha=0.78$ when
convection is suppressed. Points below (above) the curve
correspond to RBNE (BNE).} \label{fig8}
\end{figure}

Some theoretical attempts to assess the influence of
non-equipartition on segregation have been recently published.
Thus, Trujillo {\em et al.} \cite{TAH03} have derived an evolution
equation for the relative velocity of the intruders starting from
the kinetic theory proposed by Jenkins and Yoon \cite{JY02}, which
applies for weak dissipation. They use constitutive relations for
partial pressures that take into account the breakdown of energy
equipartition between the two species. However, the influence of
temperature gradients, which exist in the vibro-fluidized regime,
is neglected in Ref.\ \cite{TAH03} because it is  assumed that the
pressure and temperature are constant in the absence of the
intruder. A more refined theory has recently been provided by Brey
{\em et al.} \cite{BRM05} in the case of a single intruder in a
vibrated granular mixture under gravity. The theory displayed in
this section covers some of the aspects not accounted for in the
previous theories \cite{BRM05,JY02,TAH03} since it is based on a
kinetic theory \cite{GD02} that goes beyond the quasi-elastic
limit \cite{JY02,TAH03} and applies for arbitrary composition
$x_1$ (and so, it reduces to the results obtained in Ref.\
\cite{BRM05} when $x_1\to 0$). This allows one to assess the
influence of composition and dissipation on thermal diffusion in
bi-disperse granular gases without any restriction on the
parameter space of the system.

To determine the dependence of the coefficient $\Lambda_{12}$ on
the parameters of the mixture, we consider a non-convecting (${\bf
u}={\bf 0}$) steady state with only gradients along the vertical
direction ($z$ axis). In this case, the mass balance equation
(\ref{2.9}) yields ${\bf j}_1={\bf j}_2={\bf 0}$, while the
momentum equation (\ref{2.10}) gives
\begin{equation}
\label{9.3} \frac{\partial p}{\partial z}=-\rho g.
\end{equation}
To first order in the spatial gradients, the constitutive equation
for the mass flux $j_{1,z}$ is given by Eq.\ (\ref{4.3}), i.e.,
\begin{equation}
j_{1,z}= -\frac{m_{1}m_{2}n}{\rho } D\frac{\partial x_1}{\partial
z}-\frac{ \rho }{p}D_{p}\frac{\partial p}{\partial z}-\frac{\rho
}{T}D^{\prime }\frac{\partial T}{\partial z}, \label{9.3.1}
\end{equation}
where the susceptibility coefficient $\chi_{ij}=0$ in the
particular case of the gravitational force. The condition
$j_{1,z}=0$ yields
\begin{equation}
\label{9.4} \frac{\partial x_1}{\partial
z}=\frac{\rho^3}{m_1m_2np}\frac{D_p}{D}g-\frac{\rho^2}{m_1m_2p}\frac{D'}{D}
\frac{\partial T}{\partial z},
\end{equation}
where use has been made of Eq.\ (\ref{9.3}). Substitution of Eq.\
(\ref{9.4}) into Eq.\ (\ref{9.1}) leads to
\begin{equation}
\label{9.5} \Lambda_{12}=\frac{n\rho^2}{\rho_1\rho_2}\frac{D'-D_p
g^*}{D},
\end{equation}
where
\begin{equation}
\label{9.6} g^*\equiv \frac{\rho g}{n\left(\partial T/\partial
z\right)}<0
\end{equation}
is the reduced gravity acceleration. Since the mutual diffusion
coefficient $D$ is positive \cite{GD02,GM06}, the sign of
$\Lambda_{12}$ is determined by the sign of the quantity
$D'-D_pg^*$. This result is general since it goes beyond the
regime of density considered.
\begin{figure}
\centering
\includegraphics[width=0.7 \columnwidth,angle=0]{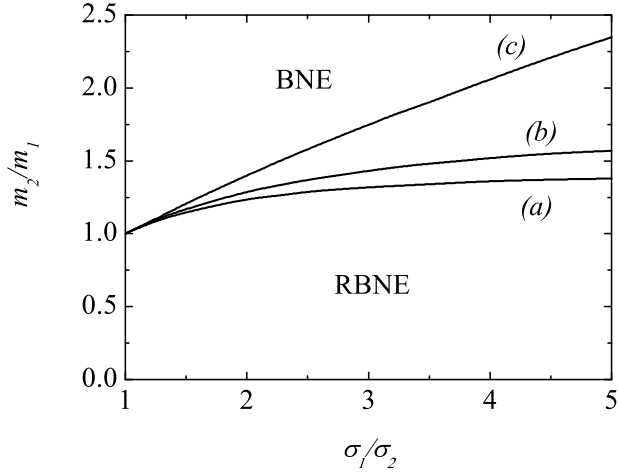}
\caption{Phase diagram for BNE/RBNE in three dimensions for
$\alpha_{ij}=0.7$ and three values of composition: (a) $x_1=0$,
(b) $x_1=0.3$, and (c) $x_1=0.7$. Points below (above) each curve
correspond to RBNE (BNE). } \label{fig9}
\end{figure}
\begin{figure}
\centering
\includegraphics[width=0.7 \columnwidth,angle=0]{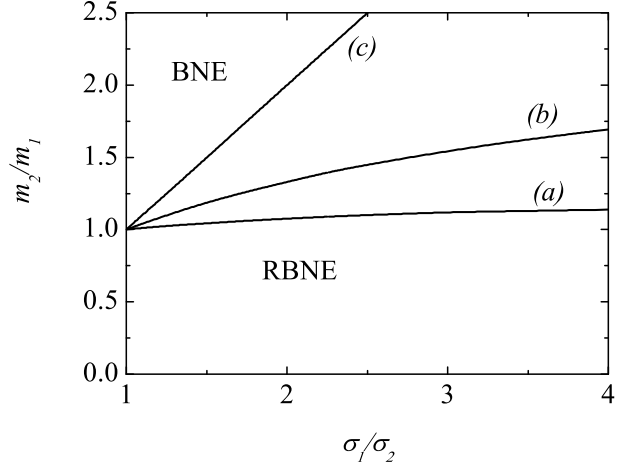}
\caption{Phase diagram for BNE/RBNE in three dimensions for
$x_1=\frac{1}{2}$ and three values of the (common) coefficient of
restitution: (a) $\alpha=0.9$, (b) $\alpha=0.8$, and (c)
$\alpha=0.5$. Points below (above) each curve correspond to RBNE
(BNE). } \label{fig10}
\end{figure}

To gain some insight into the explicit dependence of $D'$ and
$D_p$ on the parameter space of the system, one has to resort to a
kinetic theory description. For a low-density gas, the expressions
of the coefficients $D'$ and $D_p$ in the first-Sonine
approximation are given by
\begin{equation}
\label{9.7} D'=0, \quad
D_{p}=\frac{\rho_{1}p}{\rho^2\nu}\frac{x_2}{x_2+x_1\gamma}
\left(\frac{\gamma}{\mu}-1\right) ,
\end{equation}
where $\nu$ is the (positive) collision frequency \cite{GM06}
\begin{equation}
\label{9.8}
\nu=\frac{2\pi^{(d-1)/2}}{d\Gamma\left(\frac{d}{2}\right)}n\sigma_{12}^{d-1}v_0
(1+\alpha_{12})
\left(\frac{\theta_1+\theta_2}{\theta_1\theta_2}\right)^{1/2}\left(x_2\mu_{21}+
x_1\mu_{12}\right).
\end{equation}
Given that the driving stochastic term does not play a neutral
role in the transport, it must be remarked that the expressions
for the transport coefficients obtained in the driven case
slightly differ from the ones derived in the free cooling case
\cite{GD02,GM06}.

Consequently, according to Eqs.\ (\ref{9.5}) and (\ref{9.7}), the
sign of $\Lambda_{12}$ is the same as that of the pressure
diffusion coefficient $D_p$. The condition $\Lambda_{12}=0$ (or
equivalently, $D_p=0$) provides the criterion for the transition
from BNE to RBNE. Equation (\ref{9.7}) shows that the sign of
$D_p$ is determined by the value of the control parameter
\begin{equation}
\label{9.10} \theta\equiv
\frac{\gamma}{\mu}=\frac{m_2T_1}{m_1T_2}.
\end{equation}
This parameter gives the mean square velocity of the large
particles relative to that of the small particles. Thus, when
$\theta>1$ ($\theta<1$), the thermal diffusion factor is positive
(negative), which leads to BNE (RBNE). The criterion for the
transition condition from BNE to RBNE is $\theta=1$, i.e.,
\begin{equation}
\label{9.11} \frac{m_1}{m_2}=\frac{T_1}{T_2}.
\end{equation}
In the case of equal granular temperatures (energy equipartition),
$\theta\to \mu^{-1}$ and so segregation is predicted for particles
that differ in mass, no matter what their diameters may be
\cite{JY02}. It must be remarked that, due to the lack of energy
equipartition, the condition $\theta=1$ is rather complicated
since it involves all the parameter space of the system. In
particular, even when the species differ only by their respective
coefficients of restitution they also segregate when subject to a
temperature gradient. This is a novel pure effect of inelasticity
on segregation \cite{G06,SGNT06}. On the other hand, the criterion
(\ref{9.11}) for the transition BNE$\Longleftrightarrow$RBNE is
the same as the one found previously in Ref.\ \cite{TAH03} when
$\alpha_{ij}$ is close to 1 and in Ref.\ \cite{BRM05} in the
intruder limit case ($x_1 \to 0$). However, as said before, the
results obtained here are more general since they cover all the
range of the parameter space of the system.

To illustrate size segregation driven by thermal diffusion, we
consider mixtures constituted by spheres ($d=3$) of the same
material and equal total volumes of large and small particles. In
this case, $m_1/m_2=(\sigma_1/\sigma_2)^3$ and $x_2/x_1=
(\sigma_1/\sigma_2)^3$. Figure \ref{fig8} shows the phase diagram
BNE/RBNE for this kind of systems. The data points represent the
simulation results obtained by Schr\"oter et al. \cite{SUKSS06}
for $\alpha=0.78$ in agitated mixtures constituted by particles of
the same density. To the best of my knowledge, this is one of the
few experiments in which thermal diffusion has been isolated from
the remaining segregation mechanisms \cite{K04}. Our results show
that, for a given value of the coefficient of restitution, the
RBNE is dominant at small diameter ratios. However, since
non-equipartition grows with increasing diameter ratio, the system
shows a crossover to BNE at sufficiently large diameter ratios.
This behavior agrees qualitatively well with the results reported
in Ref.\ \cite{SUKSS06} at large shaking amplitudes, where thermal
diffusion becomes the relevant segregation mechanism. At a
quantitative level, we observe that the results are also
consistent with the simulation results reported in \cite{SUKSS06}
when periodic boundary conditions are used to suppress convection
since they do not observe a change back to BNE for diameter ratios
up to 3 (see red squares in Fig.\ 11 of \cite{SUKSS06}). Although
the parameter range explored in MD simulations is smaller than the
one analyzed here, one is tempted to extrapolate the simulation
data presented in Ref.\ \cite{SUKSS06} to roughly predict the
transition value of the diameter ratio at $\alpha=0.78$ (which is
the value of the coefficient of restitution considered in the
simulations). Thus, if one extrapolates from the simulation data
at the diameter ratios of $\sigma_1/\sigma_2=2$ and
$\sigma_1/\sigma_2=3$, one sees that the transition from RBNE to
BNE might be around $\sigma_1/\sigma_2=10$, which would
quantitatively agree with the results reported in Fig.\
\ref{fig8}. Figure \ref{fig8} also shows that the BNE is
completely destroyed in the quasielastic limit ($\alpha \simeq
1$).

Let us now investigate the influence of composition on
segregation. Figure \ref{fig9} shows a typical phase diagram in
the three-dimensional case for $\alpha_{ij}\equiv \alpha=0.7$ and
three different values of the mole fraction $x_1$. The lines
separate the regimes between BNE and RBNE. We observe that the
composition of the mixture has significant effects in reducing the
BNE as the concentration of larger particles increases. In
addition, for a given value of composition, the transition from
BNE to RBNE may occur following two paths: (i) along the constant
mass ratio $m_2/m_1$ with increasing size ratio
$\sigma_1/\sigma_2$, and (ii) along the constant size ratio with
increasing mass ratio $m_2/m_1$. The influence of dissipation on
the phase diagrams BNE/RBNE is illustrated in Fig.\ \ref{fig10}
for $d=3$ in the case of an equimolar mixture ($x_1=\frac{1}{2}$)
and three values of the (common) coefficient of restitution
$\alpha$. We observe that the role played by inelasticity is quite
important since the regime of RBNE increases significantly with
dissipation. Similar results are found for other values of
composition.

In summary, thermal diffusion (which is the relevant segregation
mechanism in agitated granular mixtures at large shaking
amplitudes) can been analyzed by the Boltzmann kinetic theory.
This theory is able to explain some of the experimental and/or MD
segregation results \cite{SUKSS06} observed within the range of
parameter space explored. A more quantitative comparison in the
dilute regime with MD simulations is needed to show the relevance
of the Boltzmann equation to analyze segregation driven by a
thermal gradient. As said before, comparison with MD simulations
in the tracer limit case ($x_1\to 0$) \cite{BRM05} for a dilute
gas has shown the reliability of the inelastic Boltzmann equation
to describe segregation. In this context, one expects that the
same agreement observed before in the intruder case \cite{BRM05}
is maintained when $x_1$ is different from zero.

\section{Steady States: Uniform Shear Flow}
\label{sec10}

In the preceding sections, the Navier--Stokes equations
(constitutive equations that are linear in the hydrodynamic
gradients) have been shown to be quite useful to describe
appropriately several problems in granular mixtures. However,
under some circumstances large gradients occur and more complex
constitutive equations are required. The need for more complex
constitutive equations does not signal a breakdown of
hydrodynamics \cite{TG98}, only a failure of the Navier--Stokes
approximation \cite{DB99}. Although in this case the
Chapman-Enskog method can be carried out to second order in
gradients (Burnett order), it is likely that failure of the
Navier-Stokes description signals the need for other methods to
construct the normal solution that are not based on a small
gradient expansion.

One of the most interesting problems in granular fluids is the
simple or uniform shear flow (USF) \cite{G03,C90}. As described in
Sec.\ \ref{sec5}, this state is characterized by uniform density
and temperature and a simple shear with the local velocity field
given by $u_{1,x}=u_{2,x}=ay,u_{y}=u_{z}=0$, where  $a$ is the
constant shear rate. The USF is a well-known nonequilibrium
problem widely studied, for both granular
\cite{C90,USF,WA99,AL02,CH02} and conventional \cite{GS03,Ha83}
gases. However, the nature of this state is quite different in
both systems since a steady state is achieved for granular fluids
when viscous (shear) heating is compensated for by energy
dissipation in collisions:
\begin{equation}
\label{11.1} aP_{xy} =- \frac{d}{2}  n T \zeta.
\end{equation}
This steady state is what we want to analyze in this section. The
balance equation (\ref{11.1}) shows the intrinsic connection
between the shear field and dissipation in the system. This
contrasts with the description of USF for elastic fluids where a
steady state is not possible unless an external thermostat is
introduced \cite{GS03}. Note that the hydrodynamic steady shear
flow state associated with the condition (\ref{11.1}) is
inherently beyond the scope of the Navier--Stokes or Newtonian
hydrodynamic equations \cite{SGD04}. The reason for this is the
existence of an internal mechanism, collisional cooling, that sets
the strength of the velocity gradient in the steady state. For
normal fluids, this scale is set by external sources (boundary
conditions, driving forces) that can be controlled to admit the
conditions required for Navier--Stokes hydrodynamics. In contrast,
collisional cooling is fixed by the mechanical properties of the
particles making up the fluid. This observation is significant
because it prevents the possibility of measuring the Newtonian
shear viscosity for granular fluids in the steady USF
\cite{AL02,CH02}. More generally, it provides a caution regarding
the simulation of other steady states to study Navier--Stokes
hydrodynamics when the gradients are strongly correlated to the
collisional cooling \cite{SGD04}.

From a microscopic point of view, the simple shear flow problem
becomes spatially uniform in the local Lagrangian frame moving
with the flow velocity ${\bf u}$. In this frame
\cite{GS03,LE72,DSBR86}, the velocity distribution functions adopt
the form: $f_i({\bf r},{\bf v})\to f_i({\bf V})$, where
$V_{k}=v_{k}-a_{k\ell}r_\ell$ is the peculiar velocity. Here,
$a_{k\ell}=a\delta_{kx}\delta_{\ell y}$. Under these conditions,
the set of Boltzmann kinetic equations (with ${\cal F}_i=0$) for
an isolated system reads
\begin{equation}
\label{10.2}
 -a V_{y}\frac{\partial}{\partial V_{x}}f_i({\bf
V})=\sum_{j=1}^2\;J_{ij}[{\bf V}|f_i,f_j]\;, \quad (i=1,2).
\end{equation}
The most relevant transport properties in a shear flow problem are
obtained from  the pressure tensor ${\sf P}={\sf P}_1+{\sf P}_2$,
where ${\sf P}_i$ is the  partial pressure tensor of the species
$i$ given by
\begin{equation}
\label{10.3} P_{i,k\ell}=m_i\int\; \dd{\bf V} V_k V_\ell f_i({\bf
V}).
\end{equation}

The trace of ${\sf P}_i$ defines the partial temperatures $T_i$ as
$T_i=\text{Tr}{\sf P}_i/dn_i$. As said before, these temperatures
measure the mean kinetic energy of each species. The elements of
the pressure tensor ${\sf P}_i$ can be obtained by multiplying the
Boltzmann equation (\ref{10.2}) by $m_i{\bf V}{\bf V}$ and
integrating over ${\bf V}$. The result is
\begin{equation}
\label{10.4} a_{km}P_{i,m\ell}+a_{\ell m}{P}_{i,mk}=\sum_{j=1}^2
{A}_{ij,k\ell},
\end{equation}
where we have introduced the collisional moments ${\sf A}_{ij}$ as
\begin{equation}
\label{10.5} A_{ij,k\ell}=m_i\int \dd{\bf V} V_k V_\ell
J_{ij}[{\bf V}|f_i,f_j].
\end{equation}
From Eq.\ (\ref{10.4}), in particular, one gets the balance
equation for the partial temperature  $T_i$
\begin{equation}
\label{10.6} aP_{i,xy}=-\frac{d}{2}p_i\zeta_i,
\end{equation}
where $p_i=n_iT_i$ is the partial pressure of species $i$ and
$\zeta_i$ is defined by Eq.\ (\ref{2.7.1}). According to Eq.\
(\ref{10.6}), the (steady) partial temperature in the simple shear
flow problem can be obtained by equating  the viscous heating term
$a|P_{i,xy}|$ to the collisional cooling term $(d/2)p_i\zeta_i$.

The determination of ${\sf A}_{ij}$ requires the knowledge of the
velocity distribution functions $f_i$. This is quite a formidable
task, even in the monocomponent case \cite{USF}. However, as in
the elastic case, one expects to get a good estimate of  ${\sf
A}_{ij}$ by using  Grad's approximation \cite{FK72}:
\begin{equation}
\label{10.7} f(\mathbf{V})\to
f_{i,M}(\mathbf{V})\left(1+\frac{m_i}{2T_i}C_{i,k\ell}V_k
V_\ell\right),
\end{equation}
where $f_{i,M}$ is a Maxwellian distribution at the temperature of
the species $i$, i.e.,
\begin{equation}
\label{10.8} f_{i,M}({\bf V})=n_i \left(\frac{m_i}{2\pi
T_i}\right)^{d/2}\exp\left(-\frac{m_iV^2}{2T_i}\right).
\end{equation}
As happens in the case of homogeneous states, in general the three
temperatures $T$, $T_1$, and $T_2$ are different in the inelastic
case. For this reason we choose the parameters in the Maxwellians
so that it is normalized to $n_i$ and provides the exact second
moment of $f_i$. The Maxwellians $f_{i,M}$ for the two species can
be quite different due to the temperature differences. This aspect
is essential in a two-temperature theory and has not been taken
into account in most of the previous studies
\cite{JM89,WA99,AL02}. The coefficient ${\sf C}_i$ can be
identified by requiring the moments with respect to ${\bf V}{\bf
V}$ of the trial function (\ref{10.7}) to be the same as those for
the exact distribution $f_i$. This leads to
\begin{equation}
\label{10.7.1} {\sf C}_i=\frac{{\sf P}_{i}}{p_i}-\openone
\end{equation}

With this approximation, the Boltzmann collisional moments ${\sf
A}_{ij}$ can be explicitly evaluated. The result is
\cite{MG02,G02}
\begin{eqnarray}
\label{10.9} {\sf
A}_{ij}&=&-\frac{2\pi^{(d-1)/2}}{d\Gamma(d/2)}m_i
n_in_j\mu_{ji}\sigma_{ij}^{d-1}\left(\frac{2T_i}{m_i}+\frac{2T_j}{m_j}\right)^{3/2}
(1+\alpha_{ij})\nonumber\\
& & \times
\left\{\mu_{ji}\left[\frac{T_j-T_i}{(m_j/m_i)T_i+T_j}+\frac{1-\alpha_{ij}}{2}
\right]\openone+\frac{1}{1+(m_iT_j/m_jT_i)}\right.
\nonumber\\
& & \left.\times \left[\frac{{\sf C}_i-{\sf
C}_j}{1+(m_jT_i/m_iT_j)} +\frac{d+3}{2(d+2)}\lambda_{ij}\left({\sf
C}_i+\frac{m_iT_j}{m_jT_i}{\sf C}_j\right)\right] \right\},
\end{eqnarray}
where
\begin{equation}
\label{10.10}
\lambda_{ij}=2\mu_{ji}\frac{T_j-T_i}{(m_j/m_i)T_i+T_j}+
\frac{\mu_{ji}}{d+3}(2d+3-3\alpha_{ij}).
\end{equation}
The partial cooling rates $\zeta_i$ can be easily obtained from
Eqs.\ (\ref{2.7.1}) and (\ref{10.9}).
\begin{figure}
\centering
\includegraphics[width=0.7 \columnwidth,angle=0]{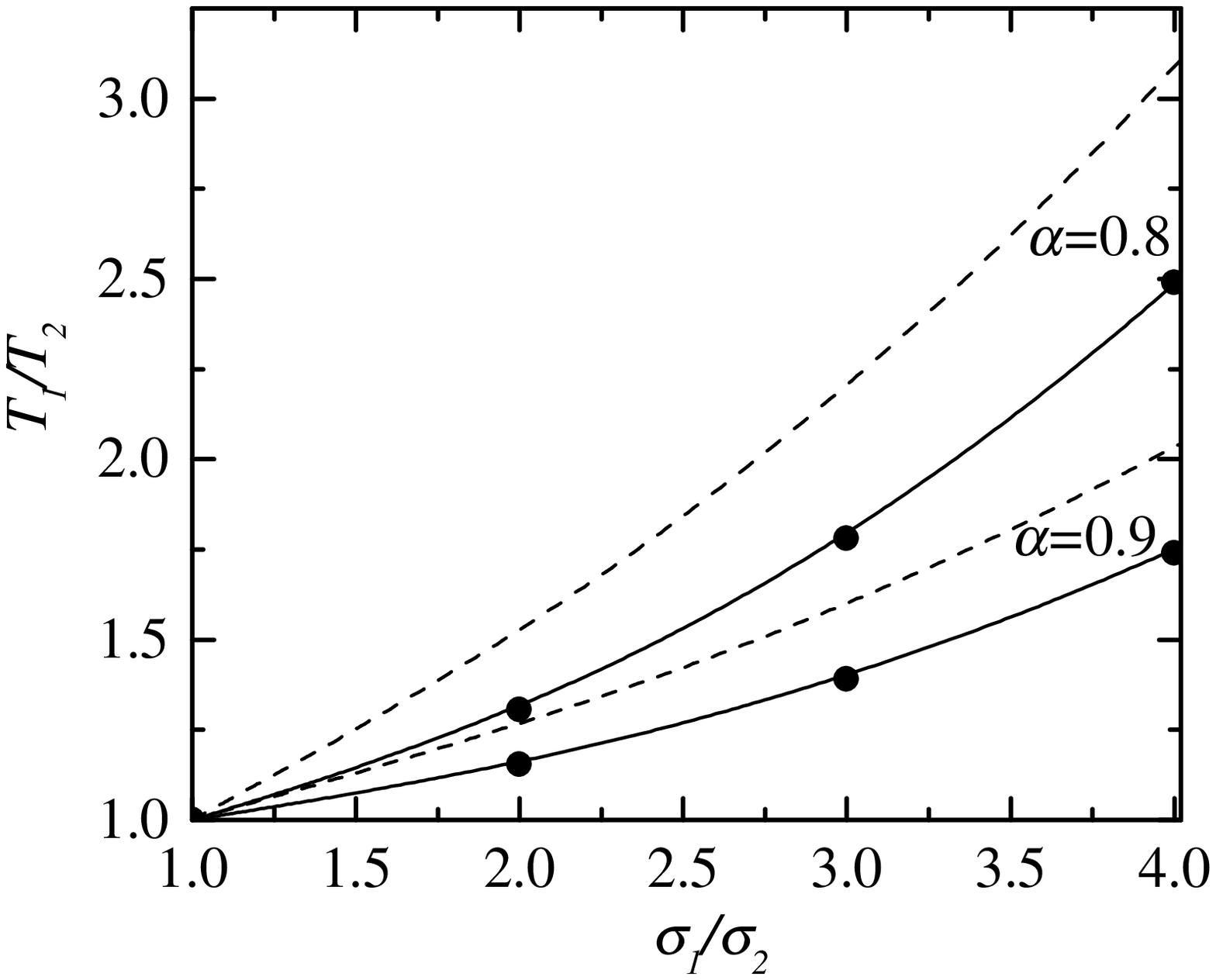}
\caption{Plot of the temperature ratio $T_1/T_2$ as a function of
the size ratio $\sigma_1/\sigma_2=(m_1/m_2)^{1/2}$ for a
two-dimensional system in the case $x_1=1/2$ and two different
values of the (common) coefficient of restitution: $\alpha=0.9$
and $\alpha=0.8$. The solid lines are the theoretical predictions
based on Grad's solution,  while the symbols refer to the DSMC
results. The dashed lines correspond to the results obtained from
the stochastic thermostat condition (\ref{3.7}).} \label{fig11}
\end{figure}

Substitution of Eq.\ (\ref{10.9}) into the set of equations
(\ref{10.4}) allows one to get the partial pressure tensor ${\sf
P}_i$ in terms of the temperature ratio $\gamma=T_1/T_2$ and the
parameters of the mixture. The temperature ratio can be obtained
from Eq.\ (\ref{10.6}) as
\begin{equation}
\label{10.11}
\gamma=\frac{x_2\zeta_2P_{1,xy}}{x_1\zeta_1P_{2,xy}}.
\end{equation}
When the expressions of ${\sf P}_i$ and $\zeta_i$ are used in Eq.\
(\ref{10.11}), one gets a {\em closed} equation for the
temperature ratio $\gamma$, that can be solved numerically. In
Fig.\ \ref{fig11} we plot $\gamma$ versus the diameter ratio
$\sigma_1/\sigma_2$ for a two-dimensional ($d=2$) granular gas
with $x_1=1/2$ and two different values of $\alpha$. The symbols
refer to the simulation data obtained from the DSMC method
\cite{GM03a}. Here, we have assumed that the disks are made of the
same material, and hence $\alpha_{ij}=\alpha$ and
$m_1/m_2=(\sigma_1/\sigma_2)^2$. The dependence of $\gamma$ on
$\sigma_1/\sigma_2$ obtained in the homogeneous steady state
driven by the stochastic thermostat (\ref{3.6}) is also included
for comparison. It is clearly seen that the kinetic theory results
based on Grad's solution agree very well with simulation data,
even for quite large values of the size ratio. In addition, the
thermostat results overestimate the simulation ones (especially
for large mass ratio), showing that the properties of the system
are not insensitive to the way at which the granular gas is
driven.
\begin{figure}
\centering
\includegraphics[width=0.7 \columnwidth,angle=0]{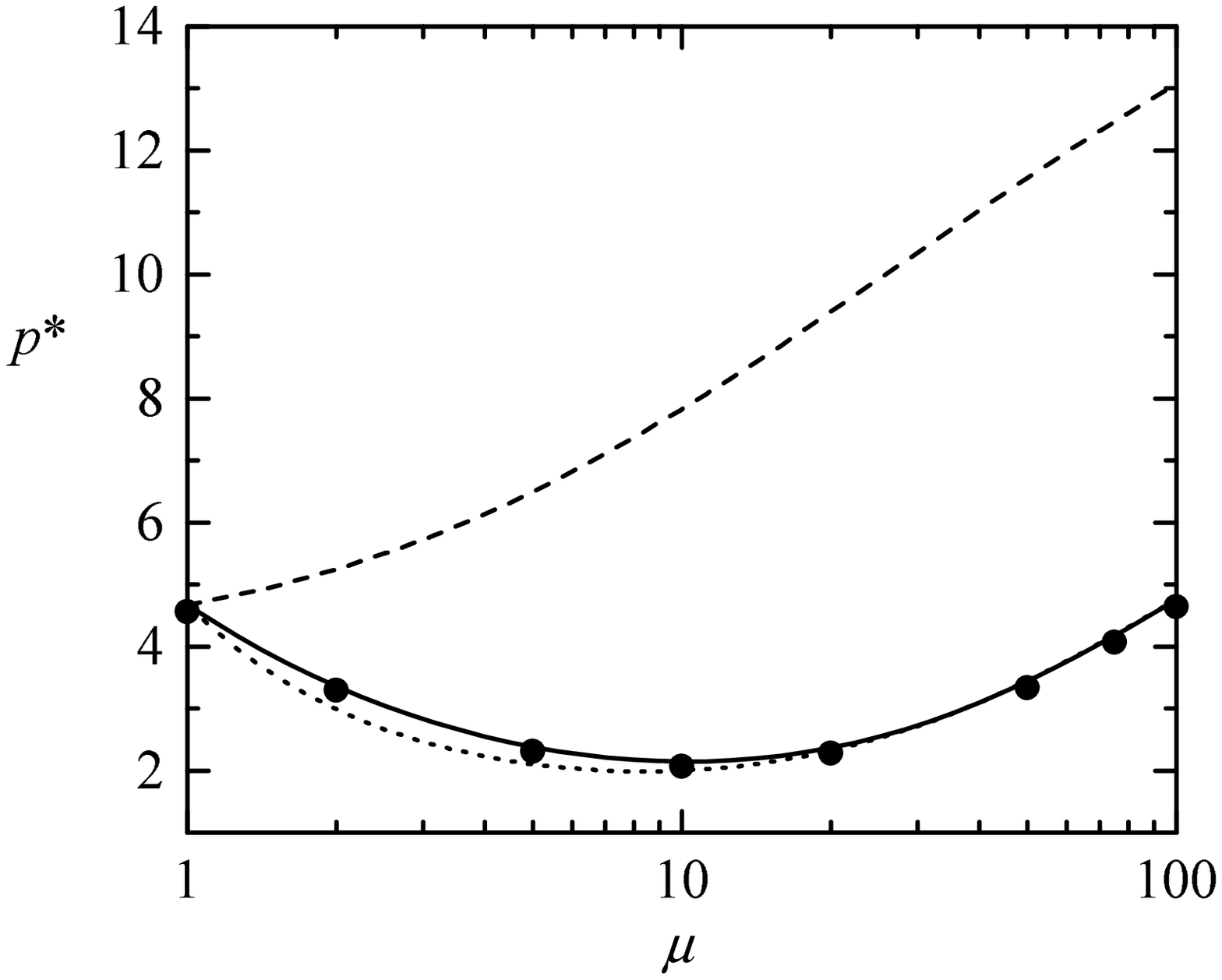}
\caption{Plot of the reduced pressure $p^*$ versus the mass ratio
$\mu=m_1/m_2$ for a two-dimensional system with
$\sigma_1=\sigma_2$, $x_1=1/2$ and $\alpha=0.9$. The solid line
corresponds to the theoretical predictions derived from Grad's
solution, the dotted line refers to the latter theory but using
the expression of $T_1/T_2$ obtained from the stochastic
thermostat condition (\ref{3.7}), and the dashed line is the
result obtained from Grad's solution by assuming the equality of
the partial temperatures ($\gamma=1$). The symbols are the DSMC
results. } \label{fig12}
\end{figure}
\begin{figure}
\centering
\includegraphics[width=0.7 \columnwidth,angle=0]{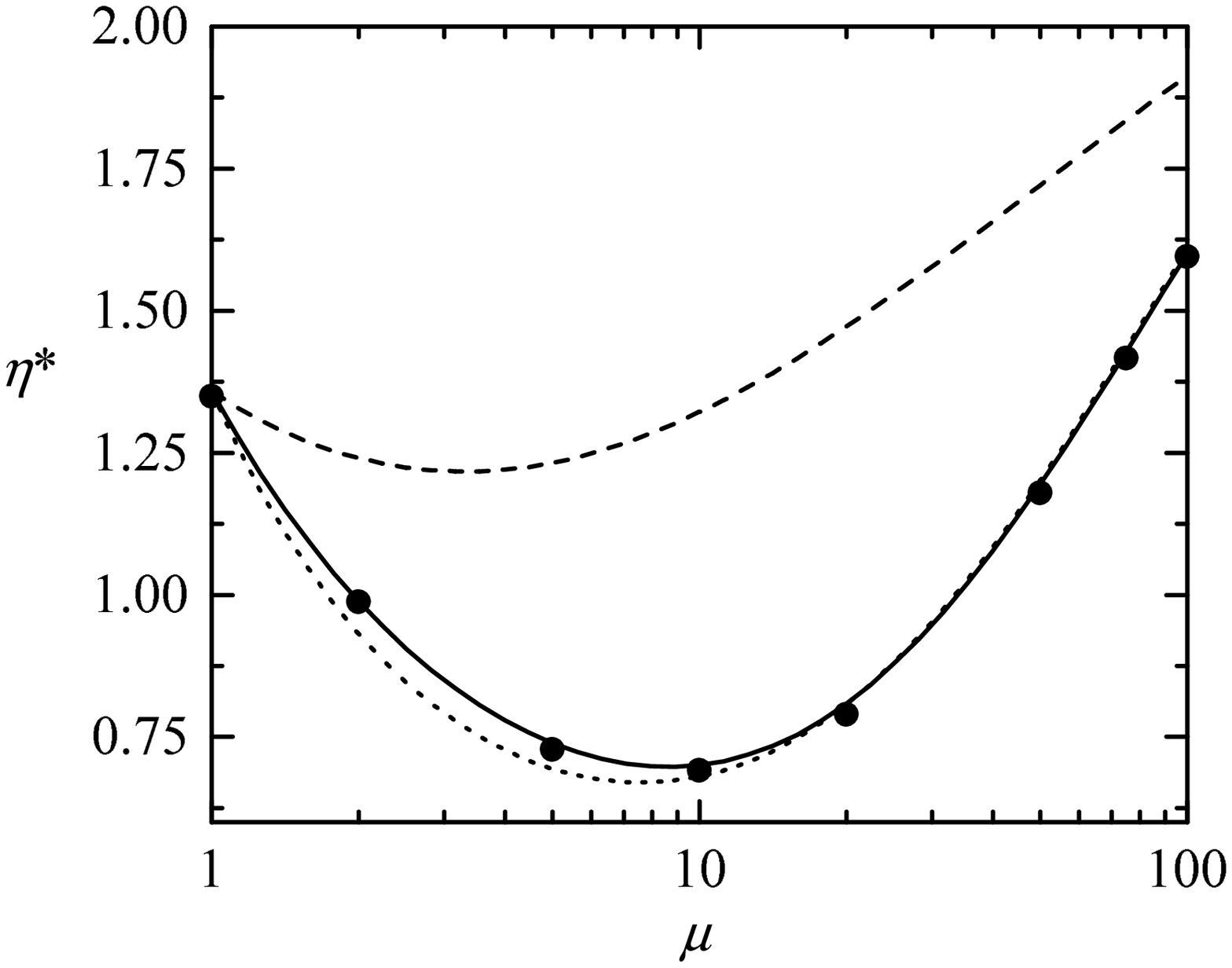}
\caption{Plot of the reduced shear viscosity $\eta^*$ versus the
mass ratio $\mu=m_1/m_2$ for a two-dimensional system with
$\sigma_1=\sigma_2$, $x_1=1/2$ and $\alpha=0.9$. The solid line
corresponds to the theoretical predictions derived from Grad's
solution, the dotted line refers to the latter theory but using
the expression of $T_1/T_2$ obtained from the stochastic
thermostat condition (\ref{3.7}), and the dashed line is the
result obtained from Grad's solution by assuming the equality of
the partial temperatures ($\gamma=1$). The symbols are the DSMC
results. } \label{fig13}
\end{figure}

Let us now consider the transport coefficients. To analyze the
rheological properties in the steady state, it is convenient to
introduce dimensionless quantities. As usual \cite{AL02}, for a
low-density gas we introduce the reduced pressure $p^*$ and the
reduced shear viscosity $\eta^*$ as
\begin{equation}
\label{10.12} p^*=\frac{p\nu^2}{\rho_1v_0^2a^2},
\end{equation}
\begin{equation}
\label{10.13} \eta^*=\frac{\eta \nu^2}{\rho_1v_0^2a},
\end{equation}
where $\eta=-P_{xy}/a$ is the non-Newtonian shear viscosity,
$P_{xy}=P_{1,xy}+P_{2,xy}$ and
$\nu=[\pi^{(d-1)/2}/\Gamma(d/2)]n\sigma_{12}^{d-1}v_0$. In Figs.\
\ref{fig12} and \ref{fig13}, we plot $p^*$ and $\eta^*$,
respectively,  as functions of the mass ratio $\mu=m_1/m_2$ for an
{\em equal-size} ($\sigma_1=\sigma_2$) binary mixture of disks
($d=2$) with $x_1=1/2$ and $\alpha=0.9$. We have also included the
predictions for $p^*$ and $\eta^*$ given by the kinetic theory but
taking the expression of $\gamma$ derived when the system is
driven by the stochastic thermostat \cite{BT02}. We observe again
in both figures an excellent agreement between the Boltzmann
theory based on Grad's solution and the DSMC results, even for
very disparate values of the mass ratio. With respect to the
influence of energy nonequipartition, Fig.\ \ref{fig12} shows that
$p^*$ presents a {\em non-monotonic} behaviour with the mass ratio
whereas the theoretical predictions with the equipartition
assumption {\em monotonically} increase with $\mu$.  In the case
of  the shear viscosity, as seen in Fig.\ \ref{fig13}, both
theories (with and without energy nonequipartition)  predict a
{\em non-monotonic} dependence of $\eta^*$ on $\mu$. However, at a
quantitative level, the influence of energy nonequipartition is
quite significant over the whole range of mass ratios considered.
The {\em non-monotonic} dependence of $p^*$ and $\eta^*$ on $\mu$
obtained here from the Boltzmann kinetic theory also agrees
qualitatively well with MD simulations carried out for bidisperse
dense systems \cite{AL02}. Thus, for instance, the minimum values
of $p^*$ and $\eta^*$ are located close to $\mu=10$ in both dilute
and dense cases. Moreover, the predictions for the transport
properties given from the present theory by taking the stochastic
thermostat expression of $\gamma$ are quite close to those
obtained from the actual value of $\gamma$, especially for large
mass ratios.

\section{Summary and Concluding Remarks}
\label{sec11}

The primary objective of this review has been to derive the
Navier--Stokes hydrodynamic equations of a binary mixture of
granular gases from the (inelastic) Boltzmann kinetic theory. The
Chapman--Enskog method \cite{CC70,FK72} is used to solve the
Boltzmann equation up to the first order in the spatial gradients
and the associated transport coefficients are given in terms of
the solutions of a set of linear integral equations. These
equations have been approximately solved by taking the leading
terms in a Sonine polynomial expansion. Comparison with controlled
numerical simulations in some idealized conditions shows quite a
good agreement between theory and simulation even for strong
dissipation. This supports the idea that the hydrodynamic
description (derived from kinetic theory) appears to be a powerful
tool for analysis and predictions of rapid flow gas dynamics of
polydisperse systems \cite{D02}.

The reference state in the Chapman--Enskog expansion has been
taken to be an exact solution of the uniform Boltzmann equation.
An interesting and important result of this solution \cite{GD99}
is that the partial temperatures (which measure the mean kinetic
energy of each species) are different. This does not mean that
there are additional degrees of freedom since the partial
temperatures can be expressed in terms of the global temperature.
This is confirmed by noting that Haff's cooling law \cite{H83} (in
the free cooling case) is the hydrodynamic mode at long
wavelengths and MD simulations confirm that the global temperature
dominates after a transient period of a few collision times
\cite{DHGD02}. In this case, only the global temperature should
appear among the hydrodynamic fields. Nevertheless, the species
temperatures play a new and interesting secondary role
\cite{GMD06}. For an ordinary (molecular) gas, there is a rapid
velocity relaxation in each fluid cell to a local equilibrium
state on the time scale of a few collisions (e.g., as illustrated
by the approach to Haff's law). Subsequently, the equilibration
among cells occurs via the hydrodynamic equations. In each cell
the species velocity distributions are characterized by the
species temperatures. These are approximately the same due to
equipartition, and the hydrodynamic relaxation occurs for the
single common temperature \cite{FK72}. A similar rapid velocity
relaxation occurs for granular gases in each small cell, but to a
universal state different from local equilibrium and one for which
equipartition no longer occurs. Hence, the species temperatures
$T_{i}$ are different from each other and from the overall
temperature $T$ of the cell. Nevertheless, the time dependence of
all temperatures (in the free cooling case) is the same in this
and subsequent states, i.e., they are proportional to the global
temperature. This implies that the species temperatures do not
provide any new dynamical degree of freedom at the hydrodynamic
stage. However, they still characterize the shape of the partial
velocity distributions and affect the quantitative averages
calculated with these distributions. The transport coefficients
for granular mixtures therefore have new quantitative effects
arising from the time independent temperature ratios for each
species \cite{GD02}. This view contrasts with some recent works
\cite{recent}, where additional equations for each species
temperature have been included among the hydrodynamic set.
However, as mentioned before, this is an unnecessary complication,
describing additional kinetics beyond hydrodynamics that is
relevant only on the time scale of a few collisions.

Another important issue discussed here has been the applicability
of the Navier--Stokes transport coefficients since their
expressions are not restricted to weak inelasticity \cite{D02}.
However, the Navier--Stokes hydrodynamic equations themselves may
or may not be limited with respect to inelasticity, depending on
the particular states studied. The Chapman--Enskog method assumes
that the relative changes of the hydrodynamic fields over
distances of the order of the mean free path are small. In the
case of ordinary fluids this can be controlled by the initial or
boundary conditions. For granular gases the situation is more
complicated since in some cases (e.g., steady states such as the
simple shear flow problem \cite{SGD04}) the boundary conditions
imply a relationship between dissipation and gradients so that
both cannot be chosen independently. In these cases, the
Navier--Stokes approximation only holds for nearly elastic
particles. However, the transport coefficients characterizing the
Navier--Stokes hydrodynamic equations are nonlinear functions of
the coefficients of restitution, regardless the applicability of
those equations.

In spite of the above cautions, the Navier--Stokes approximation
is appropriate and accurate for a wide class of flows. One group
refers to spatial perturbations of the homogeneous cooling state
(HCS) for an isolated system. Both MD and DSMC simulations
\cite{BRC99} have confirmed the dependence of the Navier--Stokes
transport coefficients on the coefficient of restitution, and
application of the Navier--Stokes hydrodynamics with these
coefficients to describe cluster formation has also been confirmed
quantitatively \cite{BRC99bis}. The same kinetic theory results
apply to driven systems as well. This is so since the reference
state is a \textit{local} HCS whose parameters vary throughout the
system to match the physical values in each cell. Examples include
application of Navier--Stokes hydrodynamics from kinetic theory to
symmetry breaking and density/temperature profiles in vertical
vibrated gases, for comparison with simulation \cite{Brey}.
Similar comparison with Navier--Stokes hydrodynamics of the latter
and of supersonic flow past a wedge in real experiments has been
given \cite{Caldera,Swinney}, showing both qualitative and
quantitative agreement. In summary, the Navier--Stokes equations
with the constitutive equations presented here remain an important
and useful description for a wide class of granular flow, although
more limited than for normal gases.

The explicit knowledge of the transport coefficients and the
cooling rate allows one to make some applications of the
Navier--Stokes hydrodynamic equations. One of them has been to
obtain the linear hydrodynamic equations for small perturbations
of the homogenous cooling state. The resulting equations exhibit a
long wavelength instability for three of the modes. This is quite
similar to the case of a monocomponent granular gas
\cite{BDKS98,BP04,G05}, and in fact the same modes are unstable
here. The additional diffusion mode for two species behaves as for
a normal fluid.

On the other hand, the constitutive equations for the mass and
heat fluxes of a granular binary mixture differ from those
obtained for ordinary fluids \cite{CC70}. This is because the
usual restrictions of irreversible thermodynamics no longer apply.
These restrictions include Onsager's reciprocal relations among
various transport coefficients and the extent to which these are
violated has also been shown here. Another application of the
Navier--Stokes equations has been to assess the violation of the
Einstein relation between the diffusion and mobility coefficients.
In the undriven case, the analysis shows that this violation is
due to three independent reasons \cite{DG01}: the absence of the
Gibbs state, the cooling of the reference state, and the
occurrence of different temperatures for the particle and
surrounding fluid. However, when the mixture is subjected to
stochastic driving, a modified Einstein relation suggested by
recent MD simulations \cite{BLP04} has also been analyzed. In this
case, the results show that the deviations of the (modified)
Einstein ratio from unity are in general very small (less than
$1\%$), in agreement with MD simulations \cite{BLP04}.

Thermal diffusion becomes the relevant segregation mechanism in
agitated granular mixtures at large shaking amplitudes. In these
conditions, the use of the Boltzmann kinetic theory for
low-density gases appears justified to understand the influence of
thermal gradient on segregation phenomena. The thermal diffusion
factor in a heated granular mixture has been explicitly evaluated
from the Chapman--Enskog solution to the Boltzmann equation. The
results show that the criterion for the transition Brazil-nut
effect $\Longleftrightarrow$reverse Brazil-nut effect is provided
by the control parameter $\theta=m_2T_1/m_1T_2$
\cite{G06,BRM05,TAH03}. Given that the energy equipartition is
broken, the condition $\theta=1$ is quite complex since it
involves all the parameters of the system: composition, masses,
sizes, and coefficients of restitution. The Boltzmann kinetic
theory results agree qualitatively well with recent MD simulations
\cite{SUKSS06} within the range of parameter space analyzed.

The hydrodynamic description also seems to be justified in the
case of steady states that are inherently beyond the scope of the
Navier--Stokes hydrodynamic equations. The reason for this
non-Newtonian behavior is the existence of an internal mechanism,
collisional cooling, that sets the scale of the spatial gradients
in the steady state. For ordinary fluids, this scale can be
externally controlled by external sources so that the conditions
for Navier--Stokes hydrodynamics apply. On the other hand, for
granular gases, collisional cooling is fixed by the mechanical
properties of the particles of the system and so the gas can
depart from the Navier--Stokes description. One well-known example
of steady states is the simple or uniform shear flow (USF).
However, in spite of the extensive prior work on USF for granular
fluids \cite{USF,AL02,CH02}, the inherent non-Newtonian character
of this state has not been conveniently taken into account. In
fact, MD simulations of steady USF have been used for granular
fluids to measure the Newtonian or Navier--Stokes shear viscosity.
The results derived here from Grad's solution and DSMC simulations
show that USF is an ideal testing ground for the study of rheology
since any choice of the shear rate and the coefficients of
restitution $\alpha_{ij}$ will provide non-Newtonian effects. It
is one of the fascinating features of granular fluids that
phenomena associated with complex fluids are more easily
accessible than for simple atomic fluids \cite{D02,G01}.

Hydrodynamics derived from hard-sphere models have found
widespread use in the description of numerous industrial processes
involving solid particles. Of particular relevance are high-speed,
gas-solid flows as found in pneumatic conveyors (of ores,
chemicals, grains, etc.) and fluidized beds (for fluid catalytic
cracking, power generation, granulation of pharmaceutical powders,
synthesis of fine chemicals like titania, etc.). Such descriptions
are now standard features of commercial and research codes. Those
codes rely upon accurate transport properties and a first order
objective is to assure this accuracy from a careful theoretical
treatment. As shown in this review, the price of this approach, in
contrast to more phenomenological approaches, is an increasing
complexity of the expressions as the systems become more complex.

The analysis carried out in this presentation has been focused on
mixtures in the dilute regime, where the collisional transfer
contributions to the transport coefficients are neglected and only
their kinetic contributions are considered. A further step is to
develop a theory for moderately dense granular mixtures. This will
provide a fundamental basis for the application of hydrodynamics
under realistic conditions. Possible extension of the present
Boltzmann kinetic theory to higher densities can be done in the
context of the Enskog kinetic equation \cite{FK72}. Preliminary
results \cite{GM03} have been restricted to the uniform shear flow
state to get directly the shear viscosity coefficient of a heated
granular mixture. The extension of this study \cite{GM03} to
states with gradients of concentration, pressure, and temperature
is somewhat intricate due to subtleties associated with the
spatial dependence of the pair correlations functions considered
in the revised Enskog theory. A future work is to extend the
results derived for moderately dense mixtures of smooth {\em
elastic} hard spheres \cite{MCK83} to inelastic collisions. This
would allow us to assess the influence of density on the different
problems addressed in this review. Of course, the precise
expressions for transport coefficients in this case will be even
more complex than for a dilute gas due to the expanded parameter
space. However, this complexity is not a problem for
implementation in a code.

As shown along this overview, granular mixtures exhibit a wide
range of interesting phenomena for which the Navier--Stokes
hydrodynamic equations can be considered as an accurate and
practical tool. However, due to their complexity, many of their
features are not fully understood. Kinetic theory and
hydrodynamics (in the broader sense) can be expected to provide
some insight into the understanding of such complex materials.

\vspace{0.75cm}

{\bf Acknowledgments}


\vspace{0.25cm}

I want to acknowledge J. W. Dufty, J. M. Montanero, and A. Santos
in their roles as collaborators and critics for much of the
material discussed here. Partial support of the Ministerio de
Ciencia y Tecnolog\'{\i}a (Spain) through Grant No.  FIS2004-01399
(partially financed by FEDER funds) and from the European
Community's Human Potential Programme HPRN-CT-2002-00307
(DYGLAGEMEM) is also acknowledged.


\begin{thebibliography} {99}


\bibitem{H83}P. K. Haff, J. Fluid Mech. {\bf 134}, 187 (1983).


\bibitem{K99}L. P. Kadanoff, Rev. Mod. Phys. {\bf 71}, 435 (1999).


\bibitem{HM86}See for instance, J.-P. Hansen and I. R. McDonald, {\em Theory of
Simple Liquids} (Academic Press, London, 1986).


\bibitem{CC70}S. Chapman and T. G. Cowling, {\em The Mathematical Theory of Nonuniform Gases}
(Cambridge University Press, Cambridge, 1970).


\bibitem{BDKS98} J. J. Brey, J. W. Dufty, C. S. Kim, and A. Santos, Phys. Rev. E {\bf 58}, 4638 (1998).


\bibitem{BC01}J. J. Brey and D. Cubero, in {\em Granular Gases}, edited by T. P{\"o}schel and S.
Luding (Lectures Notes in Physics, Springer, Vol. 564, 2001), pp.
59--78.


\bibitem{BRC99}J. J. Brey, M. J. Ruiz-Montero, and D. Cubero,
Europhys. Lett. {\bf 48}, 359 (1999).



\bibitem{BRCG00} J. J. Brey, M. J. Ruiz-Montero, D. Cubero, and R. Garc\'{\i}a-Rojo, Phys. Fluids
{\bf 12}, 876 (2000).



\bibitem{BM04}J. J. Brey and M. J. Ruiz-Montero, Phys. Rev. E {\bf 70}, 051301 (2004).


\bibitem{MSG05}J. M. Montanero, A. Santos, and V. Garz\'o, in
{\em Rarefied Gas Dynamics 24}, edited by M. Capitelli (American
Institute of Physics, Vol. 762, 2005), pp. 797--802; preprint
cond-mat/0411219.


\bibitem{D99}J. W. Dufty, J. Phys.: Condens. Matt. {\bf 12}, A47
(2000).


\bibitem{D02}J. W. Dufty, in {\em Recent Research Develop. in
Stat. Phys.} {\bf 2}, 21 (2002) and preprint cond-mat/0108444.


\bibitem{G03}I. Goldhirsch, Annu. Rev. Fluid Mech. \textbf{35}, 267 (2003).


\bibitem{JM89}J. T. Jenkins and F. Mancini, Phys. Fluids A {\bf 1}, 2050 (1989); P. Zamankhan,
Phys. Rev. E {\bf 52}, 4877 (1995); B. Arnarson and J. T. Willits,
Phys. Fluids {\bf 10}, 1324 (1998); J. T. Willits and B. Arnarson,
{\em ibid.} {\bf 11}, 3116 (1999); M. Alam, J. T. Willits, B.
Arnarson, and S. Luding, {\em ibid.} {\bf 14}, 4085 (2002); B.
Arnarson and J. T. Jenkins, {\em ibid.} {\bf 16}, 4543 (2004); D.
Serero, I. Goldhirsch, S. H. Noskowicz, and M.-L. Tan, J. Fluid
Mech. {\bf 554}, 237 (2006).


\bibitem{GD99}V. Garz\'o and J. W. Dufty, Phys. Rev. E {\bf 60}, 5706 (1999).



\bibitem{computer}See for instance, J. M. Montanero and V. Garz\'o,
Gran. Matt. {\bf 4}, 17 (2002); A. Barrat and E. Trizac, {\em
ibid.} {\bf 4}, 57 (2002); S. R. Dahl, C. M. Hrenya, V. Garz\'o,
and J. W. Dufty, Phys. Rev. E {\bf 66}, 041301 (2002); R. Pagnani,
U. M. B. Marconi, and A. Puglisi, Gran. Matt. {\bf 66}, 051304
(2002); D. Paolotti, C. Cattuto, U. M. B. Marconi, and A. Puglisi,
{\em ibid.} {\bf 5}, 75 (2003); P. Krouskop and J. Talbot, Phys.
Rev. E {\bf 68}, 021304 (2003); H. Wang, G. Jin, and Y. Ma, {\em
ibid.} {\bf 68}, 031301 (2003); J. J. Brey, M. J. Ruiz-Montero,
and F. Moreno, Phys. Rev. Lett. {\bf 95}, 098001 (2005); Phys.
Rev. E {\bf 73}, 031301 (2006); M. Schr\"oter, S. Ulrich, J.
Kreft, J. B. Swift, and H. L. Swinney, Phys. Rev. E {\bf 74},
011307 (2006).


\bibitem{exp}R. D. Wildman and D. J. Parker, Phys. Rev. Lett.
{\bf 88}, 064301 (2002); K. Feitosa and N. Menon, {\em ibid.} {\bf
88}, 198301 (2002); M. Schr\"oter, S. Ulrich, J. Kreft, J. B.
Swift, and H. L. Swinney, Phys. Rev. E {\bf 74}, 011307 (2006).


\bibitem{JM87}J. Jenkins and F. Mancini, J. Appl. Mech. {\bf 54}, 27 (1987).



\bibitem{GD02} V. Garz\'o and J. W. Dufty, Phys. Fluids {\bf 14}, 1476 (2002).


\bibitem{GMD06}V. Garz\'o, J. M. Montanero, and J. W. Dufty, Phys. Fluids {\bf
18}, 083305 (2006).



\bibitem{MG03}J. M. Montanero and V. Garz\'o, Phys. Rev. E {\bf 67}, 021308 (2003).


\bibitem{GM04}V. Garz\'o and J. M. Montanero, Phys. Rev. E {\bf 69}, 021301 (2004).



\bibitem{DG01}J. W. Dufty and V. Garz\'o, J. Stat. Phys. {\bf
105}, 723 (2001).


\bibitem{G04}V. Garz\'o, Physica A {\bf 343}, 105 (2004).




\bibitem{G06}V. Garz\'o, Europhys. Lett. {\bf 75}, 521 (2006).



\bibitem{C90}C. S. Campbell, Annu. Rev. Fluid Mech. \textbf{22}, 57 (1990).


\bibitem{USF}In the case of a monocomponent system, see for instance,
C. K. K. Lun, S. B. Savage, D. J. Jeffrey, and N. Chepurniy, J.
Fluid Mech. {\bf 140}, 223 (1984); J. T. Jenkins and M. W.
Richman, {\em ibid.} {\bf 192}, 313 (1988); C. S. Campbell, {\em
ibid.} {\bf 203}, 449 (1989); M. A. Hopkins and  H. H. Shen, {\em
ibid.} {\bf 244}, 477 (1992); P. J. Schmid and H. K. Kyt\"omaa,
{\em ibid.} {\bf 264}, 255 (1994); C. K. K. Lun and A. A. Bent,
{\em ibid.} {\bf 258}, 335 (1994); I. Goldhirsch and M. L. Tan,
Phys. Fluids {\bf 8}, 1752 (1996); N. Sela, I. Goldhirsch, and S.
H. Noskowicz, {\em ibid.} {\bf 8}, 2337 (1997); J. J. Brey, M. J.
Ruiz-Montero, and F. Moreno, Phys. Rev. E {\bf 55}, 2846 (1997);
C.-S. Chou and M. W. Richman, Physica A {\bf 259}, 430 (1998);
C.-S. Chou {\em ibid.} {\bf 287}, 127 (2000); {\bf 290}, 341
(2001); J. M. Montanero, V. Garz\'o, A. Santos, and J. J. Brey, J.
Fluid Mech. {\bf 389}, 391 (1999); A. Astillero and A. Santos,
Phys. Rev. E {\bf 72}, 031309 (2005).



\bibitem{GS03}V. Garz\'o and A. Santos, {\em Kinetic Theory of Gases in Shear Flows.
Nonlinear Transport} (Kluwer Academic, Dordrecht, 2003).



\bibitem{SGD04}A. Santos, V. Garz\'o, and J. W. Dufty,
Phys. Rev. E {\bf 69}, 061303 (2004).



\bibitem{GS95}A. Goldshtein and M. Shapiro, J. Fluid Mech. {\bf 282}, 75 (1995).


\bibitem{BDS97}J. J. Brey, J. W. Dufty, and A. Santos, J. Stat. Phys. {\bf 87},
1051 (1997).


\bibitem{NE98}T. P. C. van Noije and M. H. Ernst, Gran. Matt. {\bf 1}, 57 (1998).


\bibitem{MG02bis}J. M. Montanero and V. Garz\'o, Gran. Matt. {\bf 4}, 17 (2002).


\bibitem{DHGD02}S. R. Dahl, C. M. Hrenya, V. Garz\'o, and J. W. Dufty, Phys. Rev. E {\bf 66}, 041301 (2002).


\bibitem{EM90}D. J. Evans and G. P. Morriss, {\em Statistical Mechanics of Nonequilibrium Liquids}
(Academic Press, London, 1990).


\bibitem{H91}W. G. Hoover, {\em Computational Statistical Mechanics} (Elsevier, Amsterdam, 1991).


\bibitem{MS00}J. M. Montanero and A. Santos, Gran. Matt. {\bf 2}, 53 (2000).




\bibitem{WM96}D. R. M. Williams and F. C. McKintosh, Phys. Rev. E {\bf 54}, R9 (1996).


\bibitem{HBB00}C. Henrique, G. Batrouni, and D. Bideau, Phys. Rev.
E {\bf 63}, 011304 (2000).


\bibitem{BT02}A. Barrat and E. Trizac, Gran. Matt. {\bf 4}, 52
(2002).


\bibitem{SUKSS06}M. Schr\"oter, S. Ulrich, J. Kreft,
J. B. Swift, and H. L. Swinney, Phys. Rev. E {\bf 74}, 011307
(2006).


\bibitem{BRM05}J. J. Brey, M. J. Ruiz-Montero, and F. Moreno, Phys. Rev. Lett.
{\bf 95}, 098001 (2005); Phys. Rev. E {\bf 73}, 031301 (2006).


\bibitem{FK72}J. Ferziger and H. Kaper, {\em Mathematical Theory
of Transport Processes in Gases} (North-Holland, Amsterdam, 1972).


\bibitem{GM06}V. Garz\'o and J. M. Montanero, preprint
cond-mat/0604552.


\bibitem{GD99b}
V. Garz\'o and  J.W. Dufty, Phys. Rev. E \textbf{59}, 5895 (1999).


\bibitem{L05}J. F. Lutsko, Phys. Rev. E {\bf 72}, 021306 (2005).


\bibitem{Bird}G. A. Bird, {\em Molecular Gas Dynamics and the Direct Simulation Monte Carlo of Gas Flows}
(Clarendon, Oxford, 1994).


\bibitem{M89}J. A. McLennan, {\em Introduction to Nonequilibrium Statistical Mechanics}
(Prentice-Hall, New Jersey, 1989).



\bibitem{SD06}A. Santos and J. W. Dufty, Phys. Rev. Lett. {\bf
97}, 058001 (2006).



\bibitem{MC84}M. L\'opez de Haro and E. G. D. Cohen, J. Chem. Phys.
{\bf 80}, 408 (1984).


\bibitem{GM02}V. Garz\'o and J. M. Montanero, Physica A {\bf 313},
336 (2002).


\bibitem{BLP04}A. Barrat, V. Loreto, and A. Puglisi, Physica A
{\bf 334}, 513 (2004).


\bibitem{GM84} S. R. de Groot and P. Mazur, \emph{Nonequilibrium
Thermodynamics} (Dover, New York, 1984).



\bibitem{BP04}N. Brilliantov and T. P{\"o}schel, {\em Kinetic
Theory of Granular Gases} (Oxford University Press, Oxford, 2004).


\bibitem{GZ93} I. Goldhirsch and G. Zanetti, Phys. Rev. Lett. {\bf 70}, 1619
(1993); I. Goldhirsch, M.L. Tan, and G. Zanetti, J. Sci. Comput.
{\bf 8}, 1 (1993).


\bibitem{LH99}S. Luding and H.J. Herrmann, Chaos {\bf 9}, 673
(1999).


\bibitem{G05} V. Garz\'o, Phys. Rev. E \textbf{72}, 021106 (2005).


\bibitem{RL77}P. R\'esibois and M. de Leener, {\em Classical
Kinetic Theory of Fluids} (John Wiley, NY, 1977).


\bibitem{BY80}J. P. Boon and S. Yip, {\em Molecular Hydrodynamics}
(Dover, NY, 1980).


\bibitem{BRC99bis}J. J. Brey, M. J. Ruiz-Montero, and D. Cubero,
Phys. Rev. E {\bf 60}, 3150 (1999).


\bibitem{K04}See, for instance, A. Kudrolli, Rep. Prog. Phys. {\bf 67}, 209 (2004).


\bibitem{KCM83}J. M. Kincaid, E. G. D. Cohen, and M. L\'opez de
Haro, J. Chem. Phys. {\bf 86}, 963 (1983).


\bibitem{HQL01}D. C. Hong, P. V. Quinn, and S. Luding, Phys. Rev. Lett. {\bf
86}, 3423 (2001).


\bibitem{JY02}J. T. Jenkins and D. K. Yoon, Phys. Rev. Lett. {\bf
88}, 194301 (2002); D. K. Yoon and J. T. Jenkins, Phys. Fluids
{\bf 18}, 073303 (2006).



\bibitem{BEKR03}A. P. J. Breu, H. M. Ensner, C. A. Kruelle, and
I. Rehberg, Phys. Rev. Lett. {\bf 90}, 014302 (2003).


\bibitem{horizontal}T. Schautz, R. Brito,
C. A. Kruelle, and I. Rehberg, Phys. Rev. Lett. {\bf 95}, 028001
(2005) and references therein.



\bibitem{TAH03}L. Trujillo, M. Alam, and H. J. Herrmann, Europhys.
Lett. {\bf 64}, 190 (2003).


\bibitem{SGNT06} D.
Serero, I. Goldhirsch, S. H. Noskowicz, and M.-L. Tan, J. Fluid
Mech. {\bf 554}, 237 (2006).


\bibitem{TG98}M. L. Tan and I. Goldhirsch, Phys. Rev. Lett. {\bf
81}, 3022 (1998).


\bibitem{DB99}J. W. Dufty and J. J. Brey, Phys. Rev. Lett. {\bf
82}, 4566 (1999).



\bibitem{WA99}J. T. Willits and B. Arnarson, Phys. Fluids {\bf 11},
3116 (1999); M. Alam, J. Willits, B. Arnarson, and S. Luding, {\em
ibid.} {\bf 14}, 4085 (2002).


\bibitem{AL02}M. Alam and S. Luding, Gran. Matt. {\bf 4}, 139
(2002); J. Fluid Mech. {\bf 476}, 69 (2003); Phys. Fluids {\bf
17}, 063303 (2005).


\bibitem{CH02}M. Alam and C. M. Hrenya, Phys. Rev. E {\bf 63},
0613018 (2001); R. Clelland and C. M. Hrenya, {\em ibid.} {\bf
65}, 031301 (2002); S. R. Dahl, R. Clelland, and C. M. Hrenya,
Phys. Fluids {\bf 14}, 1972 (2002); S. R. Dahl, R. Clelland, and
C. M. Hrenya, Powder Tech., {\bf 138}, 7 (2003); S. R. Dahl and C.
M. Hrenya, Phys. Fluids {\bf 16}, 1 (2004); H. Iddir, H.
Arastoopour, and C. M. Hrenya, Powder Tech. {\bf 151}, 117 (2005).




\bibitem{Ha83}See for instance, {\em Nonlinear Fluid Behavior}, edited by H. J. M.
Hanley (North-Holland, Amsterdam, 1983).



\bibitem{LE72}A. W. Lees and S. F. Edwards, J. Phys. C {\bf 5}, 1921 (1972).


\bibitem{DSBR86}J. W. Dufty, A. Santos, J. J. Brey, and R. F. Rodr\'{\i}guez,
Phys. Rev. A {\bf 33}, 459 (1986).




\bibitem{MG02}J. M. Montanero and V. Garz\'o, Physica A {\bf 310}, 17 (2002).



\bibitem{G02}V. Garz\'o, Phys. Rev. E \textbf{66}, 021308 (2002).


\bibitem{GM03a}V. Garz\'o and J. M. Montanero, Gran. Matt. {\bf 5},
165 (2003).


\bibitem{recent} L. Huilin, D. Gidaspow, and E. Manger,
Phys. Rev. E \textbf{64}, 061301 (2001); M.F. Ramahan, J. Naser,
and P. J. Witt, Powder Tech. \textbf{138}, 82 (2003); J. E.
Galvin, S. R. Dahl, and C. M. Hrenya, J. Fluid Mech. \textbf{528},
207 (2005).





\bibitem{Brey} J. J. Brey, M. J. Ruiz-Montero, F. Moreno, and R.
Garc\'{\i}a-Rojo, Phys. Rev. E \textbf{63}, 061305 (2001); {\em
ibid.} \textbf{65}, 061302 (2002)



\bibitem{Caldera} X. Yang, C. Huan, D. Candela, R. W. Mair, and R. L.
Walsworth, Phys. Rev. Lett. {\bf 88}, 044301 (2002); C. Huan, X.
Yang, D. Candela, R. W. Mair, and R. L. Walsworth, Phys. Rev. E
\textbf{69}, 041302 (2004).



\bibitem{Swinney} C. Bizon, M. D. Shattuck, J. B. Swift, and Harry L.
Swinney, Phys. Rev. E \textbf{60}, 4340 (1999); E. C. Rericha, C.
Bizon, M. D. Shattuck, and H. L. Swinney, Phys. Rev. Lett.
\textbf{88}, 014302 (2002).


\bibitem{G01}I. Goldhirsch, in {\em Granular Gases}, edited by T. P{\"o}schel and S.
Luding (Lectures Notes in Physics, Springer, Vol. 564, 2001), pp.
79--99.


\bibitem{GM03}V. Garz\'o and J. M. Montanero, Phys. Rev. E {\bf 68}, 041302 (2003).


\bibitem{MCK83}M. L\'opez de Haro, E. G. D. Cohen, and J. M.
Kincaid, J. Chem. Phys. {\bf 78}, 2746 (1983).





\end{thebibliography}
\end{document}